\documentclass[lettersize,journal]{IEEEtran}

\usepackage{url}
\usepackage{bbm}
\usepackage{cite}
\usepackage{array}
\usepackage{xcolor}
\usepackage{comment}
\usepackage{flushend}
\usepackage{makecell}
\usepackage{booktabs}
\usepackage{textcomp}
\usepackage{colortbl}
\usepackage{verbatim}
\usepackage{graphicx}
\usepackage{textcomp}
\usepackage{stfloats}
\usepackage{algorithm}
\usepackage{algorithmic}
\usepackage{amsmath, amssymb, amsfonts}
\usepackage[caption=false,font=normalsize,labelfont=sf,textfont=sf]{subfig}
\usepackage[colorlinks,linkcolor=blue,anchorcolor=black,citecolor=blue]{hyperref}

\newtheorem{lemma}{Lemma}

\newtheorem{theorem}{Theorem}

\newtheorem{definition}{Definition}
\newtheorem{proposition}{Proposition}

\begin{document}

\title{RIS-Assisted Joint Sensing and Communications via Fractionally Constrained Fractional Programming}

\author{Yiming~Liu,~\IEEEmembership{Graduate~Student~Member,~IEEE,}
		Kareem~M.~Attiah,~\IEEEmembership{Member,~IEEE,}
		and
		Wei~Yu,~\IEEEmembership{Fellow,~IEEE}
		
		\thanks{
			This work was supported by the Natural Sciences and Engineering Research Council (NSERC) via a Discovery Grant and via the Canada Research Chairs program.
			An earlier version of this paper was presented in part at the IEEE Global Communications Conference (GLOBECOM), December 2024 \cite{Liu_GLOBECOM2024}.
			\textit{(Corresponding author: Wei Yu.)}	
			
			The authors are with The Edward S. Rogers Sr. Department of Electrical and Computer Engineering, University of Toronto, Toronto, Ontario M5S 3G4,
			Canada (e-mails: eceym.liu@mail.utoronto.ca; kattiah@ece.utoronto.ca; weiyu@ece.utoronto.ca).
			}
		
}



\maketitle

\begin{abstract}
	
This paper studies an uplink dual-functional sensing and communication system aided by a reconfigurable intelligent surface (RIS), whose reflection pattern is configured to trade-off sensing and communication functionalities.
Specifically, the Bayesian Cram\'{e}r-Rao lower bound (BCRLB) for estimating the azimuth angle of a sensing user is minimized while ensuring the signal-to-interference-plus-noise ratio constraints for communication users.
We show that this problem can be formulated as a novel fractionally constrained fractional programming (FCFP) problem.
To deal with this nontrivial optimization problem, we extend a quadratic transform technique, originally proposed to handle optimization problems containing fractional structures only in objectives, to the scenario where the constraints also include ratios.
First, we consider the case where the fading coefficient is known.
Using the quadratic transform, the FCFP problem can be turned into a sequence of subproblems that are convex except for the constant-modulus constraints which can be tackled using a penalty-based approach.
To further reduce the computational complexity, we leverage the constant-modulus conditions and propose a novel linear transform.
This new transform enables the FCFP problem to be turned into a sequence of linear programming (LP) subproblems, which can be solved efficiently.
Then, we consider the case where the fading coefficient is unknown.
A modified BCRLB is used to make the problem more tractable, and the proposed quadratic transform based algorithm is used to solve the problem.
Numerical results unveil nontrivial and effective reflection patterns that can be synthesized by the RIS to facilitate both communication and sensing functionalities.

\end{abstract}


\section{Introduction}

\IEEEPARstart{W}{ith} the increasing demands on positioning information and for sensing functionality in the beyond-fifth-generation (B5G) and sixth-generation (6G) wireless networks, integrated sensing and communications (ISAC) is viewed as a promising use case for future networks (see, e.g., \cite{9737357} and the references therein), particularly when operating at the millimeter-wave (mmWave) and terahertz (THz) bands \cite{9540344, 9606831}.
However, signal propagation characteristics at these high frequencies are challenging for both sensing and communications due to the penetrating pathloss.
In particular, due to obstacles that are often present in practical scenarios, sensing and/or communications users are often located in regions with no direct line-of-sight (LoS) paths from the base station (BS).
However, the non-line-of-sight (NLoS) paths are often too weak to ensure good performance.

To address this issue, reconfigurable intelligent surface (RIS) \cite{8741198}, also known as intelligent reflecting surface (IRS) \cite{8811733}, has emerged as a viable and promising solution for undertaking sensing and communication tasks in regions with no direct LoS paths.
Specifically, RIS is a cost-effective planar reflector capable of altering the phases of incident electromagnetic (EM) waves with very low power consumption.
It has the ability to generate controllable and desired reflection patterns, and to create new LoS paths, in order to enable joint sensing and communications in practical deployment scenarios with many obstructions \cite{10243495}.

This paper investigates an \emph{uplink} RIS-assisted joint sensing and communications problem, where the direct paths between the users and the BS are blocked and the reflection patterns of the RIS need to be optimally configured in order to serve both sensing and communication purposes.
Specifically, the sensing user actively transmits pilots to the BS through the RIS, allowing the BS to estimate the channel characteristics of the sensing user based on the received pilots. 
At the same time, the communication users transmit data symbols to the BS through the RIS.
In this paper, we separate the sensing and communication users spatially via BS and RIS beamforming.
This is possible, because the numbers of BS antennas and RIS elements typically exceed the number of users.

Unlike conventional radar-based downlink sensing, pilot-based uplink sensing enables sensing users to autonomously determine the timing and necessity of initiating sensing services;
they only actively transmit pilots when sensing services are needed.
In addition, uplink sensing also helps conserve BS resources and reduce the burden on the BS.

There are numerous prior works investigating joint sensing and communications with/without the assistance of RIS.
For example, the authors in \cite{8386661} design the beampatterns for a dual-functional radar-communication system in order to achieve a flexible trade-off between two functionalities.
To facilitate the MIMO radar in utilizing its full degree-of-freedom (DoF), the authors in \cite{9124713} jointly design the individual radar and communication waveforms to optimize both the radar beampatterns and the signal-to-interference-and-noise ratio (SINR) for the communication users.
In \cite{9724202}, the authors consider an RIS-enabled target estimation, where dedicated sensors are installed at the RIS to estimate the directions of nearby targets.
In \cite{9361184, 9454375, 9732186}, the authors investigate RIS-enabled target detection, where the RIS beamformer is designed to maximize the target detection probability.
In \cite{9769997,10364735,9364358}, the authors study RIS-assisted ISAC systems, where the beamformers are optimized to maximize the SNR for sensing while satisfying certain quality-of-service (QoS) constraints.
Apart from the aforementioned studies that focus on downlink scenarios, several initial investigations are conducted in uplink ISAC systems, e.g., systems architecture design \cite{8999605, 9618653}, performance analysis \cite{9800940, 10608079, 10542219}, and beamformer design \cite{10472418, 10158711, 10557567}.
In \cite{10274660}, the authors explore the joint beamforming and power allocation for an uplink RIS-assisted ISAC system.
These works investigate the potential of uplink ISAC systems and provide an initial attempt at addressing the challenging joint signal detection and sensing problem.

The sensing metrics adopted in the aforementioned works, e.g., SNR maximization and beampatterns matching, are all based on heuristics.
This is in contrast to the Cram\'{e}r-Rao lower bound (CRLB) metric, which is more directly connected to the sensing performance, because it provides a lower bound on the mean-squared error (MSE) for parameter estimation among all unbiased estimators.
In this realm, \cite{9652071} adopts the CRLB to evaluate the target estimation performance in a system without RIS.
In \cite{10138058}, the CRLB for an RIS-aided wireless sensing system is minimized, although communication functions are not integrated into the system.
In \cite{10440056}, the authors further incorporate the communication function based on the prior work in \cite{10138058}.
The authors in \cite{9591331, 9416177} propose to utilize RIS in uplink ISAC systems to reduce the interference and the CRLB, respectively.

It should be noted that the standard CRLB depends on the actual values of the parameters to be estimated (such as the angles-of-arrival and the channel fading coefficients), which are unknown in practice.
A more practical framework is to assume a prior distribution for the parameters to be estimated and to employ the Bayesian CRLB (BCRLB) (e.g., \cite{6541985}) as a more suitable sensing metric.

In this paper, we investigate an RIS-assisted dual-functional uplink sensing and communication system, where the beamformer at the RIS is designed to balance the tradeoff between the two functions.
We deal with a specific problem of minimizing the BCRLB for an angle estimation problem with one sensing user and multiple communication users under SINR constraints. 
Different from radar-based downlink ISAC, where the sensing signals are received at the BS and the communication signals are received at the user side, in the uplink case, both the sensing and the communication signals are transmitted from the users through the RIS and received at the BS. 
Because of the interference among the user signals, we show in this paper that the uplink joint sensing and communications problem leads to a novel fractionally constrained fractional programming (FCFP) problem formulation with extra RIS amplitude constraints.

To solve this nontrivial optimization problem, we extend a quadratic transform technique, previously proposed to handle optimization problems for which the objective function has a fractional structure \cite{8314727}, to the scenario where the constraints also contain ratio terms.
We first consider the problem in the case of known fading coefficient.
We show that the quadratic transform allows the original problem to be transformed into a sequence of subproblems which are convex except for the extra constant-modulus constraints introduced by the RIS.
A penalty-based method is then utilized to handle the constant-modulus constraints.
To further reduce the complexity, we leverage the constant-modulus condition and propose a new transform called the constant-modulus linear transform.
This new technique allows the FCFP problem to be transformed into a sequence of linear programming (LP) subproblems, which can be solved efficiently.

We also consider the case of unknown fading coefficient.
In this scenario, a more complicated problem is formulated with a modified BCRLB.
The quadratic transform technique and the penalty-based method can also be used to solve this problem.
Numerical results demonstrate the effectiveness of the proposed algorithms and show that the optimization can produce nontrivial RIS reflection patterns that facilitate both sensing and communication functionalities at the same time.

\subsubsection*{Paper Organization}

The rest of this paper is organized as follows.
Section~\ref{Section_02} introduces the system and signal models.
Section~\ref{Section_03} defines the performance metrics and formulates the beamforming problem for joint sensing and communications.
Section~\ref{Section_04} proposes a penalty and quadratic transform based algorithm to tackle the problem.
Section~\ref{Section_05} proposes a linear transform to reduce the computational complexity.
Section~\ref{Section_06} extends the problem to the case involving unknown fading coefficients.
Section~\ref{Section_07} presents simulation results and interprets the obtained solutions. 
Section~\ref{Section_08} concludes the paper.

\subsubsection*{Notations}

Scalars, vectors, and matrices are represented by lowercase letters, bold lowercase letters, and bold uppercase letters, respectively.
The transpose and Hermitian transpose of matrices are represented by $\left( \cdot \right)^{\mathsf{T}}$ and $\left( \cdot \right)^{\dagger}$, respectively.
The complex Gaussian distribution is represented by $\mathcal{CN} \left( \cdot, \cdot \right)$, and the expectation is denoted by $\mathbb{E} \left[ \cdot \right]$.
The operator $\mathrm{vec} \left( \cdot \right)$ stacks the columns of a matrix into a column vector, and $\mathrm{vec}^{-1} \left( \cdot \right)$ denotes its inverse operation.

\section{System and Signal Model}
\label{Section_02}

\subsection{Channel Model}

\begin{figure}[!t]
	\centering
	\includegraphics[width = 1 \linewidth]{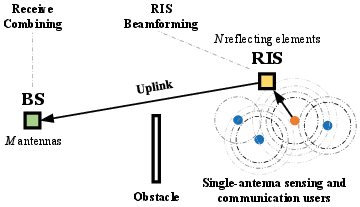}
	\caption{An uplink RIS-assisted joint sensing and communication system.}
	\label{Fig_01}
\end{figure}

Consider an uplink RIS-assisted dual-functional sensing and communication system, where a BS with $M$ antennas intends to estimate the azimuth angle from a single-antenna sensing user to the RIS, while simultaneously receiving communication signals from $K$ single-antenna users, as shown in Fig.~\ref{Fig_01}.

The RIS performs a linear mapping of the incident signals, which can be modeled by a multiplication with an equivalent diagonal phase-shift matrix $\mathop{\mathrm{diag}} \left\lbrace \mathbf{x} \right\rbrace$.
The vector $\mathbf{x}$ represents the reflection coefficients~of the RIS and is expressed as
\begin{align}
	\mathbf{x}
	\triangleq
	\left[
		e^{j \theta_1},
		e^{j \theta_2},
		\ldots,
		e^{j \theta_{N}}
	\right]^{\mathsf{T}},
\end{align}
where $\theta_n \in [-\pi, \pi)$ denotes the phase shift of the $n$-th element of the RIS, $n = 1, \ldots, N$, and $N = N_{\mathsf{row}} \times N_{\mathsf{col}}$ is the number of reflecting elements in the RIS. 
In this paper, we design $\mathbf{x}$ to enable joint sensing and communications.
Note that in this model of RIS beamforming, the optimization variable $\mathbf{x}$ has~a constant-modulus constraint, i.e., $\left| x_n \right| = 1$, $\forall n$, which may be a challenging constraint to handle.

We assume that all the direct channels between the users and the BS are obstructed.
Consequently, the radio signals can only propagate through the RIS.
The ability to help establish links in situations where the direct channels are obstructed is one of the crucial advantages of RIS.
Note that the methodology developed in this paper can be readily extended to scenarios that include direct channels.

For simplicity, the channel from each user to the RIS is assumed to contain only one dominant LoS path\footnote{
	The mmWave channel can often be modeled by a dominant LoS path \cite{6847111}. 
	As RIS is often deployed in close proximity to the user side, we only consider the dominant LoS path in the user-RIS link.
	However, this work is applicable more generally to the case where the user-RIS links are modeled as a Rician fading channel.}.
Furthermore, the array response of the RIS is assumed to be a function only of the azimuth angle of the LoS path.
Specifically, the sensing user is assumed to be located in a two-dimensional plane with a fixed elevation angle, so that the overall array response of~the uniform rectangular array of RIS elements depends only on the sensing user's azimuth angle-of-arrival $\eta$ as follows:
\begin{align}
	\mathbf{v}(\eta) = 
	\left[
		e^{j \tau \cos \left( \eta \right) v \left( 1 \right)},
		\ldots, 
		e^{j \tau \cos \left( \eta \right) v \left( N \right)}
	\right]^{\mathsf{T}},
\end{align}
where $\tau = {2 \pi d} / {\omega}$, and $\omega$ represents the carrier wavelength, $d$ denotes the spacing between two adjacent reflecting elements (typically at half wavelength), and $v \left( n \right) \triangleq \mathop{\mathrm{mod}} \left( n - 1, N_{\mathsf{col}} \right)$.
In this case, the cascaded reflection channel from the sensing user to the BS through the RIS becomes
\begin{align} \label{los_channel}
	\mathbf{u}
	=
	\mathbf{G} \mathop{\mathrm{diag}} \{ \mathbf{x} \} \alpha \mathbf{v}(\eta)
	=
	\alpha \mathbf{G} \mathop{\mathrm{diag}} \{ \mathbf{v}(\eta) \} \mathbf{x}
	\triangleq
	\alpha \mathbf{U}(\eta) \mathbf{x},
\end{align}
where $\alpha$ denotes the complex fading coefficient of the channel from the sensing user to the RIS, and $\mathbf{G}$ denotes the channel from the RIS to the BS, which is shared across all the sensing and communication users.
The sensing task considered in this paper is for the BS to estimate the azimuth angle $\eta$.

For the communication users, let $\phi_k$ represent the azimuth angle of the $k$-th user towards the RIS.
The cascaded reflection channel of this user can similarly be expressed as
\begin{align}
	\mathbf{h}_k
	= 
	\beta_k \mathbf{G} \mathop{\mathrm{diag}} \{ \mathbf{v}(\phi_k) \} \mathbf{x}
	\triangleq 
	\mathbf{H}_k \mathbf{x},
\end{align}
where $\beta_k$ denotes the complex fading coefficient of the channel between this communication user and the RIS.
In this paper, we assume that all the communication channels $\mathbf{H}_k$, as well as the shared RIS-BS channel $\mathbf{G}$, are perfectly known at the BS.
This assumption is reasonable due to the fixed positions of the BS and the RIS, as well as the availability of reliable channel estimation techniques (see, e.g., \cite{9722893} and the references therein).
The communications objective is to satisfy certain SINR constraint for each of the communication users.

\subsection{Signal Model for Joint Sensing and Communications}

This paper investigates the following joint sensing and communications signaling strategy.
Within the channel coherence time, represented by $T$, the $K$ communication users transmit symbols $s_1^{\mathsf{c}}[t], \ldots, s_K^{\mathsf{c}}[t]$, respectively, for $t = 1, \ldots, T$.
These symbols are modeled as independent and identically distributed (i.i.d.) Gaussian random variables following $\mathcal{CN} \left( 0, p_k \right)$, where $p_k$ is the fixed transmit power of the $k$-th communication user, for $k = 1, \ldots, K$.
In the meanwhile, the sensing user transmits a sequence of known pilots $s^{\mathsf{o}}[t]$ of fixed power $p$ synchronously to the BS, for $t = 1, \ldots, T$.

Based on the channel model established in the previous sub-section, the signal received at the BS can be expressed as
\begin{align} \label{signal_model}
	\mathbf{y}[t]
	= 
	\alpha \mathbf{U}(\eta) \mathbf{x}[t] s^{\mathsf{o}}[t] 
	+ 
	\sum_{k=1}^{K} \mathbf{H}_k \mathbf{x}[t] s_k^{\mathsf{c}}[t] 
	+ 
	\mathbf{n}[t], 
\end{align}
for $t = 1,\ldots, T$, where $\mathbf{n}[t]$ represents the additive Gaussian noise distributed as $\mathcal{CN} \left( \mathbf{0}, \sigma_n^2 \mathbf{I} \right)$.

This paper aims to design the RIS reflection coefficients $\mathbf{x}[t]$ in order to enable joint sensing and communications.
This is done by adaptively setting $\mathbf{x}[t]$ in each symbol period $t$.

Specifically, the sensing operation is assumed to take place in one symbol period at a time.
The optimization process aims to design the RIS coefficients to minimize the estimation error for sensing $\eta$ in the next symbol period, while ensuring communication performance is maintained.
The adaptation of $\mathbf{x}[t]$ is based on the effective channels $\mathbf{H}_k$ of the communication users, the channel model $\alpha \mathbf{U}(\eta)$ for the sensing user, as well as the prior knowledge of unknown parameters $\eta$ and $\alpha$.

This paper assumes that the numbers of the BS antennas and the RIS reflecting elements exceed the number of served users, and relies on beamforming at the RIS and at the BS to spatially separate the sensing and communication users.
We assume that no interference cancellation is performed at the BS.
Although interference cancellation may further improve the performance (see, e.g., \cite{10663294, YuarXiv}), it comes at a cost of increased delay and complexity, since it requires the completion of one task (e.g., estimating angles or detecting data symbols) before starting of the next task.
In contrast, separating all the users in the spatial domain allows the sensing and all the communication tasks to be performed simultaneously.

\section{Sensing and Communication Performance Metrics and Problem Formulation} \label{Section_03}

In this section, we establish the performance metrics of both sensing and communications.
Specifically, We use the BCRLB as a metric of sensing performance, while adopting the SINR as a measure of communication quality.
Based on the two established metrics, we formulate the optimization problem for the task of joint sensing and communications.
For simplicity, the fading coefficient $\alpha$ is assumed to be known for now.
The scenario where $\alpha$ is unknown, so that both $\alpha$ and $\eta$ are required to be estimated, is considered later in this paper.

\subsection{Performance Metric for Sensing}

We first derive the classic CRLB, then present its Bayesian version, i.e., the BCRLB, as a metric for sensing.
Let $\hat{\eta}$ be an unbiased estimator of $\eta$ over one symbol period.
According to the Cram\'{e}r-Rao theorem \cite{kay1993fundamentals}, the mean square error (MSE) of any unbiased estimator for $\eta$ is bounded from below as
\begin{align}
	\mathop{\mathbb{E}} \left| \hat{\eta} - \eta \right|^2
	\geq
	\mathsf{CRLB} \left( \eta \right) 
	\triangleq
	\frac{1}{\mathsf{FI} \left( \eta \right)},
\end{align}
where $\mathsf{FI} \left( \eta \right)$ is the Fisher information, defined as 
\begin{equation}
	\mathsf{FI} \left( \eta \right)
	\triangleq
	\mathop{\mathbb{E}} \!
	\left[ 
		\left( \frac{\partial \ln \mathcal{L} \left( \mathbf{y}; \eta \right)}{\mathop{\partial} \eta} \right)^2
	\right],
\end{equation}
and $\mathcal{L} \left( \mathbf{y}; \eta \right)$ is the likelihood function of $\mathbf{y}$ as function of $\eta$.
Since the pilot signal from the sensing user is known, and both the communication signals and the noise are modeled as Gaussian random vectors, $\mathcal{L} \left( \mathbf{y}; \eta \right)$ is the probability density function of a Gaussian distribution with the mean and the covariance matrix given as below: 
\begin{align}
	\mathsf{mean} \left( \mathbf{y} \mathop{|} \eta \right)
	& = 
	\alpha \mathbf{U}(\eta) \mathbf{x} s^{\mathsf{o}},  \label{mean}  \\
	\mathsf{Cov} \left( \mathbf{y} \mathop{|} \eta \right)
	& = 
	\sum_{k=1}^{K} 
	p_k
	\left( \mathbf{H}_{k} \mathbf{x} \right) 
	\left( \mathbf{H}_{k} \mathbf{x} \right)^{\dagger}
	+ 
	\sigma_n^2 \mathbf{I}
	\triangleq 
	\mathbf{\Sigma}^{\mathsf{o}}(\mathbf{x}).  \label{cov}
\end{align}
Then, the Fisher information $\mathsf{FI} \left( \eta \right)$ can be expressed as \cite{9500437}
\begin{align} \label{FI}
	\mathsf{FI} \left( \eta \right)
	=
	2 \hspace{1pt} p
	\left( \alpha \dot{\mathbf{U}}(\eta) \mathbf{x} \right)^{\dagger}
	\left( \mathbf{\Sigma}^{\mathsf{o}}(\mathbf{x}) \right)^{-1}
	\left( \alpha \dot{\mathbf{U}}(\eta) \mathbf{x} \right) ,
\end{align}
where $\dot{\mathbf{U}}(\eta)$ denotes the derivative of $\mathbf{U}(\eta)$ with respect to $\eta$.

Note here that we drop the time index $t$. 
This is because we consider the sensing operation over one symbol period only. 
The derivation can be easily extended to cases where sensing takes place over multiple symbol periods.

One can observe that $\mathsf{FI} \left( \eta \right)$ has a fractional structure with respect to $\mathbf{x}$, attributed to the fact that both the sensing and communication signals propagate through the RIS and interact with each other at the BS.
In particular, the communication signals are regarded as interference for sensing, since sensing and communication are assumed to take place simultaneously without interference cancellation.

CRLB has been widely used to characterize the performance for parameter estimation;
it is an alternative metric to the MSE when the MSE is difficult to compute.
However, it can be seen from \eqref{FI} that the computation of CRLB depends on the exact values of the parameters to be estimated, which are unknown.
Consequently, we cannot directly employ this classic CRLB.

To address this issue, we assume a prior distribution of the parameters of interest and adopt the Bayesian version BCRLB to evaluate sensing performance \cite{van2002optimum, 1523369}.
The prior distribution can be updated based on successive observations.

Following the derivations in \cite{10584278, 1703855}, the BCRLB of $\eta$ with a prior distribution $q \left( \eta \right)$ can be expressed as follows:
\begin{align} \label{BCRLB}
	\mathsf{BCRLB} \left( \eta \right)
	& =
	\frac{1}{
		\mathop{\mathbb{E}_{q \left( \eta \right)}}\! \left[ \mathsf{FI} \left( \eta \right) \right]
		+ 
		\mathsf{FIP} \left( \eta \right)
	},
\end{align}
where $\mathsf{FIP} \left( \eta \right)$ is the Fisher information of the prior itself and is independent of the optimization variable $\mathbf{x}$.
One can observe from \eqref{BCRLB} that minimizing the BCRLB over $\mathbf{x}$ is equivalent to maximizing the following new metric:
\begin{align} \label{11}
	\mathcal{A}(\mathbf{x})
	& \triangleq
	\frac{\mathop{\mathbb{E}_{q \left( \eta \right)}}\! \left[ \mathsf{FI} \left( \eta \right) \right]}{2 \hspace{1pt} p \hspace{-1pt} \left| \alpha \right|^2}  \notag \\
	& =
	\mathop{\mathbb{E}_{q \left( \eta \right)}}\!
	\left[ 
	\left( \dot{\mathbf{U}} ( \eta ) \mathbf{x} \right)^{\dagger}
	\left( \mathbf{\Sigma}^{\mathsf{o}}(\mathbf{x}) \right)^{-1}
	\left( \dot{\mathbf{U}} ( \eta ) \mathbf{x} \right)
	\right].
\end{align}
However, computing $\mathcal{A}$ still requires an integration over $q \left( \eta \right)$. 
The theorem below extracts $\mathbf{x}$ from the integral.

\begin{theorem}
	\label{thm:second_moment}
	Consider a second moment matrix defined as
	\begin{equation} \label{U}
		\dot{\mathbf{R}}
		= 
		\mathop{\mathbb{E}_{q \left( \eta \right)}} \!
		\left[ 
			\mathop{\mathrm{vec}} \! \left( \dot{\mathbf{U}}(\eta) \right)  
			\mathrm{\mathop{vec}}^{\dagger} \! \left( \dot{\mathbf{U}}(\eta) \right) 
		\right].
	\end{equation}
	Let the eigendecomposition of this matrix be
	\begin{equation} \label{eigen}
		\dot{\mathbf{R}}
		= 
		\sum_{r=1}^{R} 
		\kappa_r \mathbf{r}_r \mathbf{r}_r^{\dagger}.
	\end{equation}
	Then, $\mathcal{A}(\mathbf{x})$ can be equivalently rewritten as 
	\begin{equation} \label{obj_Approx}
		\mathcal{A}(\mathbf{x})
		= 
		\sum_{r=1}^{R}
		\kappa_r
		\left( \mathbf{R}_{r} \mathbf{x} \right)^{\dagger} 
		\left( \mathbf{\Sigma}^{\mathsf{o}}(\mathbf{x}) \right)^{-1}
		\left( \mathbf{R}_{r} \mathbf{x} \right),
	\end{equation}
	where $\kappa_r > 0$, and $\mathbf{R}_r = \mathrm{\mathop{vec}}^{-1} \! \left( \mathbf{r}_r \right)$.
\end{theorem}

\begin{IEEEproof}
	The proof is given in Appendix~\ref{app:theorem_1}.
\end{IEEEproof}

Based on Theorem~\ref{thm:second_moment}, the metric $\mathcal{A}$ for estimating $\eta$ can be transformed into a weighted sum of multi-dimensional ratios in $\mathbf{x}$ as \eqref{obj_Approx}.
Note that the standard CRLB is a special case of the above, where $\eta$ is deterministic and the prior distribution $q \left( \eta \right)$ has zero variance.
In this case, $\dot{\mathbf{R}}$ is rank-one, and \eqref{obj_Approx} reduces to \eqref{FI}.

\subsection{Performance Metric for Communications}

For the communication users, we employ a unit-norm linear combiner $\mathbf{w}_k$ at the BS to perform detection and demodulation of the data symbols for the $k$-th user.
Since no interference cancellation is performed at the BS, the pilot signals from the sensing user are regarded as interference to the communication users.
Furthermore, since the azimuth angle $\eta$ of the sensing user is unknown, we opt to use the expected power of the pilot signal over its prior distribution $q \left( \eta \right)$ as the interference power in the SINR expression.
Although strictly speaking, the interference due to an actual realization of $\eta$ may lead to a higher interference level than the expected interference power, the use of expected interference power leads to a more tractable problem formulation.

With these assumptions, the SINR of the $k$-th communication user can now be expressed as
\begin{align} \label{sinrfisrt}
	& \gamma_k (\mathbf{x}) =  \notag  \\
	& \quad 
	\frac{
		p_k \left| \mathbf{w}_k^{\dagger} \mathbf{H}_{k} \mathbf{x} \right|^2}{
		p \mathop{\mathbb{E}_{q \left( \eta \right)}}\! \left[ \left| \alpha \mathbf{w}_k^{\dagger} \mathbf{U}(\eta) \mathbf{x} \right|^2 \right]
		+ \mathop{\displaystyle \sum_{j \neq k}}
		p_{j} \left| \mathbf{w}_k^{\dagger} \mathbf{H}_{j} \mathbf{x} \right|^2
		+ \sigma_n^2}. 
\end{align}
For a fixed RIS reflection coefficient $\mathbf{x}$, optimizing the receive combining vector $\mathbf{w}_k$ to maximize the SINR is a generalized Rayleigh quotient problem: 
\begin{equation}
	\begin{aligned}
		\mathop{\mathrm{maximize}}_{\mathbf{w}_k} \;\;
		\frac{p_k \mathbf{w}_{k}^{\dagger} \left( \left( \mathbf{H}_{k} \mathbf{x} \right) \left( \mathbf{H}_{k} \mathbf{x} \right)^{\dagger} \right) \mathbf{w}_{k}}{\mathbf{w}_{k}^{\dagger} \mathbf{\Sigma}^{\mathsf{c}}_{k} (\mathbf{x}) \mathbf{w}_{k}},
	\end{aligned}
\end{equation}
where $\mathbf{\Sigma}^{\mathsf{c}}_{k} (\mathbf{x})$ is the covariance matrix of the combined interference and noise
\begin{multline} \label{20}
	\mathbf{\Sigma}^{\mathsf{c}}_{k} (\mathbf{x})
	=
	\sum_{j \neq k} p_{j} \left( \mathbf{H}_{j} \mathbf{x} \right) \left( \mathbf{H}_{j} \mathbf{x} \right)^{\dagger}  \\
	+ 
	p \mathop{\mathbb{E}_{q \left( \eta \right)}} \! \left[ \left( \alpha \mathbf{U}(\eta) \mathbf{x} \right) \left( \alpha \mathbf{U}(\eta) \mathbf{x} \right)^{\dagger} \right]
	+ 
	\sigma_n^2 \mathbf{I}.
\end{multline}
The expectation term in \eqref{20} can be rewritten as
\begin{multline} \label{reWr}
	\mathop{\mathbb{E}_{q \left( \eta \right)}} \!
	\left[ \left( \alpha \mathbf{U}(\eta) \mathbf{x} \right) \left( \alpha \mathbf{U}(\eta) \mathbf{x} \right)^{\dagger} \right] = \\
	\mathrm{\mathop{vec}}^{-1} \Big( 
	\mathop{\mathbb{E}_{q \left( \eta \right)}} \!
	\left[ \left( \alpha \mathbf{U}(\eta) \right)^{*} \otimes \left( \alpha \mathbf{U}(\eta) \right) \right]  \mathrm{\mathop{vec}} \! \left( \mathbf{x} \mathbf{x}^{\dagger} \right) \! \Big),  
\end{multline}
so $\mathbf{x}$ can be taken out of the expectation operation.
The optimal solution $\mathbf{w}_{k}^{\star}$ can then be found in closed-form as
\begin{equation} \label{update wm}
	\mathbf{w}_{k}^{\star}
	= \left( \left( \mathbf{\Sigma}^{\mathsf{c}}_{k}(\mathbf{x}) \right)^{-1} \mathbf{H}_{k} \mathbf{x} \right) 
	\mathop{/}
	\left\| \left( \mathbf{\Sigma}^{\mathsf{c}}_{k}(\mathbf{x}) \right)^{-1} \mathbf{H}_{k} \mathbf{x} \right\|_2.
\end{equation}
With this optimal receive combining vector, the SINR in \eqref{sinrfisrt} can be rewritten as a multi-dimensional ratio as
\begin{equation} \label{msinr}
	\gamma_k (\mathbf{x})
	= 
	p_{k} 
	\left( \mathbf{H}_{k} \mathbf{x} \right)^{\dagger} 
	\left( \mathbf{\Sigma}^{\mathsf{c}}_{k}(\mathbf{x}) \right)^{-1} 
	\left( \mathbf{H}_{k} \mathbf{x} \right).
\end{equation}
In this paper, we consider the joint design of receive combiners and RIS reflection coefficients to ensure that a minimal SINR threshold is satisfied for each user.

\subsection{Problem Formulation}

Based on the performance metrics discussed in the previous two subsections, the problem of optimally designing the RIS reflection coefficients for joint sensing and communications can now be formulated as
\begin{subequations}
	\label{prob:joint_sensing}
	\begin{align}
		\textbf{{(P1):}} \quad
		\mathop{\mathrm{maximize}}_{\mathbf{x}} \;\;
		& \;
		\mathcal{A}(\mathbf{x})  \\
		\mathop{\mathrm{subject\;to}} \;\;
		&
		\frac{\gamma_k(\mathbf{x})}{p_k} \geq \Gamma_k, \ \forall k, \\
		&
		\left| x_n \right| = 1, \ \forall n,
	\end{align}
\end{subequations}
where $\Gamma_k = \frac{\Gamma}{p_k}$ and $\Gamma$ represents the SINR threshold for all the communication users.
The threshold value affects a trade-off between the two functionalities. 
For example, when $\Gamma$ equals zero, the problem reduces to a sensing-only problem.

The above problem formulation assumes a prior $q \left( \eta \right)$ and optimizes $\mathbf{x}$ for the next symbol period.
This process can be repeated in a sequential fashion within the channel coherence time, so that the RIS reflecting coefficients $\mathbf{x}$ can be adjusted for the subsequent symbols, while the prior distribution $q \left( \eta \right)$ is updated to account for the observations made over time.

We note here that when $\mathbf{x}$ is adjusted within the coherence time, the communication receivers can update their knowledge of the effective channels using the model $\mathbf{h}_k = \mathbf{H}_k \mathbf{x}$, so that the
symbol detector can be updated accordingly without having to re-estimate the effective channel.
The SINR expressions remain valid for the new effective channel.

As already mentioned, in this paper, the sensing operation is assumed to take place over one symbol period at a time.
But the proposed problem formulation can also be readily extended to the case where each sensing operation takes place over multiple symbol periods by modifying the BCRLB expression to be over multi\-ple symbol periods.

The rest of this paper focuses on solving the problem (P1).
Observe that this problem is an FCFP problem, because it has sum-of-ratios as the objective function (i.e., \eqref{obj_Approx}) and it has fractional constraints (i.e., \eqref{msinr}).
In the next two sections, we present two different algorithms to tackle problem (P1).

\section{Quadratic Transform Based Algorithm for RIS Reflection Coefficient Design} 
\label{Section_04}

In this section, we consider utilizing the quadratic transform technique and the penalty-based method to solve problem (P1).
We first modify the quadratic transform technique to make it applicable to problems with fractional constraints.
Then, we utilize the penalty-based algorithm to deal with the constant-modulus constraints.

\subsection{Quadratic Transform for FCFP}

We begin by defining a standard multi-dimensional ratio.
\begin{definition}
	A standard multi-dimensional ratio is a function $\mathbb{C}^N \to \mathbb{R}_{+}$ of the form
	\begin{align} \label{ratio}
		\mathsf{f} (\mathbf{x})
		\triangleq
		\left( \mathbf{A} \mathbf{x} \right)^{\dagger}
		\left( \mathbf{D}(\mathbf{x}) \right)^{-1}
		\left( \mathbf{A} \mathbf{x} \right),
	\end{align}
	where the denominator matrix $\mathbf{D}(\mathbf{x})$ has the form
	\begin{equation}
		\mathbf{D}(\mathbf{x})
		\triangleq
		\sum_{m}
		\varrho_m
		\left( \mathbf{B}_{m} \mathbf{x} \right)
		\left( \mathbf{B}_{m} \mathbf{x} \right)^{\dagger}
		+
		\mathbf{C},
	\end{equation}
	the weight $\varrho_m > 0$, and $\mathbf{C}$ is positive definite.
\end{definition}

In a study of optimization problems involving ratio objective functions \cite{8314727}, the following quadratic transform is proposed as a technique to decouple the numerator and the denominator of a ratio.
This result is summarized in the following lemma.

\begin{lemma}
	\label{lem:quadratic_transform}
	A lower bound for a standard multi-dimensional ratio function $\mathsf{f}(\mathbf{x})$ in \eqref{ratio} is the following:
	\begin{align} \label{Minor1}
		\mathsf{f}(\mathbf{x})
		\geq 
		\tilde{\mathsf{f}}(\mathbf{x}, \boldsymbol{\lambda})
		\triangleq
		2 \mathop{\mathfrak{Re}} \! \left\lbrace \boldsymbol{\lambda}^{\dagger} \mathbf{A} \mathbf{x} \right\rbrace
		- 
		\boldsymbol{\lambda}^{\dagger} \mathbf{D}(\mathbf{x}) \boldsymbol{\lambda},
		\ \forall \mathbf{x}, \boldsymbol{\lambda},
	\end{align}
	with the equality achieved at
	\begin{equation} \label{op_aux1}
		\boldsymbol{\lambda}^{\star}
		= 
		\left( \mathbf{D}(\mathbf{x}) \right)^{-1} \mathbf{A} \mathbf{x}.
	\end{equation}
\end{lemma}

In this paper, we extend this quadratic transform technique to FCFP problems, in which the ratio functions also appear in the constraints, as shown in the following theorem.

\begin{theorem}
	\label{thm:FCFP}
	Consider a multi-dimensional-function-of-ratios maximization problem
	\begin{subequations} \label{q}
		\begin{align}
			\mathop{\mathrm{maximize}}_{\mathbf{x} \in \mathcal{X}} \
			&\
			\mathsf{h}^{\mathsf{o}} \! \left( 
				\mathsf{f}^{\mathsf{o}}_1(\mathbf{x}), 
				\ldots, 
				\mathsf{f}^{\mathsf{o}}_I(\mathbf{x}) \right)  \\
			\mathop{\mathrm{subject\;to}} \
			&\
			\mathsf{h}^{\mathsf{c}}_k \! \left( 
				\mathsf{f}^{\mathsf{c}}_{k,1}(\mathbf{x}), 
				\ldots, 
				\mathsf{f}^{\mathsf{c}}_{k,J}(\mathbf{x}) \right) \geq c_k, \ \forall k, \label{Rayc}
		\end{align}
	\end{subequations}
	where $\mathsf{f}^{\mathsf{o}}_i (\mathbf{x})$ and $\mathsf{f}^{\mathsf{c}}_{k,j} (\mathbf{x})$ are standard multi-dimensional ratios of the form \eqref{ratio}, and $\mathcal{X}$ is a nonempty feasible set. 

Suppose that the following conditions hold: 
\begin{itemize}
\item	The outer function in the objective, $\mathsf{h}^{\mathsf{o}}(\cdot) \! : \mathcal{H}^{\mathsf{o}} \rightarrow \mathbb{R}$, is nondecreasing in each component on a domain $\mathcal{H}^{\mathsf{o}}$.
	Similarly, the outer function in each constraint, $\mathsf{h}_k^{\mathsf{c}}(\cdot) \! : \mathcal{H}_k^{\mathsf{c}} \rightarrow \mathbb{R}$, is also nondecreasing in each component on a domain $\mathcal{H}_k^{\mathsf{c}}$.
\item The image of $\mathcal{X}$ under $\mathbf{f}^{\mathsf{o}}(\cdot) = \left( \mathsf{f}^{\mathsf{o}}_1(\cdot), \ldots, \mathsf{f}^{\mathsf{o}}_I(\cdot) \right)$ is a subset of $\mathcal{H}^{\mathsf{o}}$, i.e.,
	\begin{align} \label{condition_01}
		\left\lbrace \mathbf{f}^{\mathsf{o}}(\mathbf{x}) \mathop{|} \mathbf{x} \in \mathcal{X} \right\rbrace \subseteq \mathcal{H}^{\mathsf{o}},
	\end{align}
	and the image of $\mathcal{X}$ under $\mathbf{f}_k^{\mathsf{c}}(\cdot) = \left( \mathsf{f}^{\mathsf{c}}_{k,1}(\cdot), \ldots, \mathsf{f}^{\mathsf{c}}_{k,J}(\cdot) \right)$ is a subset of $\mathcal{H}_k^{\mathsf{c}}$, i.e.,
	\begin{align} \label{condition_02}
		\left\lbrace \mathbf{f}_k^{\mathsf{c}}(\mathbf{x}) \mathop{|} \mathbf{x} \in \mathcal{X} \right\rbrace \subseteq \mathcal{H}_k^{\mathsf{c}}, \ \forall k.
	\end{align}
	\end{itemize}
	Then, problem~\eqref{q} is equivalent to
	\begin{subequations} \label{equivalent Ray}
		\begin{align}
			\mathop{\mathrm{maximize}}_{\mathbf{x} \in \mathcal{X}, \boldsymbol{\lambda}^{\mathsf{o}}_i, \boldsymbol{\lambda}^{\mathsf{c}}_j} \
			&\;
			\mathsf{h}^{\mathsf{o}} \! \left( \tilde{\mathsf{f}}^{\mathsf{o}}_1 (\mathbf{x}, \boldsymbol{\lambda}^{\mathsf{o}}_1), \ldots, \tilde{\mathsf{f}}^{\mathsf{o}}_I (\mathbf{x}, \boldsymbol{\lambda}^{\mathsf{o}}_I) \right)  \\
			\mathop{\mathrm{subject\;to}} \
			&\;
			\mathsf{h}^{\mathsf{c}}_k \! \left( \tilde{\mathsf{f}}^{\mathsf{c}}_{k,1} (\mathbf{x}, \boldsymbol{\lambda}^{\mathsf{c}}_{k,1}), \ldots, \tilde{\mathsf{f}}^{\mathsf{c}}_{k,J} (\mathbf{x}, \boldsymbol{\lambda}^{\mathsf{c}}_{k,J}) \right) \geq c_k, \ \forall k, \label{29b} \\
			&
			\left( \tilde{\mathsf{f}}^{\mathsf{o}}_1 (\mathbf{x}, \boldsymbol{\lambda}^{\mathsf{o}}_1), \ldots, \tilde{\mathsf{f}}^{\mathsf{o}}_I (\mathbf{x}, \boldsymbol{\lambda}^{\mathsf{o}}_I) \right) \in \mathcal{H}^{\mathsf{o}},  \label{range_01}  \\
			&
			\left( \tilde{\mathsf{f}}^{\mathsf{c}}_{k,1} (\mathbf{x}, \boldsymbol{\lambda}^{\mathsf{c}}_{k,1}), \ldots, \tilde{\mathsf{f}}^{\mathsf{c}}_{k,J} (\mathbf{x}, \boldsymbol{\lambda}^{\mathsf{c}}_{k,J}) \right) \in \mathcal{H}_k^{\mathsf{c}}, \ \forall k,  \label{range_02}
		\end{align}
	\end{subequations}
	where the transformed functions $\tilde{\mathsf{f}}^{\mathsf{o}}_i (\mathbf{x}, \boldsymbol{\lambda}^{\mathsf{o}}_i)$ and $\tilde{\mathsf{f}}^{\mathsf{c}}_{k,j} (\mathbf{x}, \boldsymbol{\lambda}^{\mathsf{c}}_{k,j})$ are quadratic functions defined in \eqref{Minor1}. 
	For each feasible $\mathbf{x}$, the optimal	$\boldsymbol{\lambda}^{\mathsf{o}}_i$ and $\boldsymbol{\lambda}^{\mathsf{c}}_{k,j}$ are functions of $\mathbf{x}$ in the form of \eqref{op_aux1} with their respective $\mathbf{D}(\mathbf{x})$ and $\mathbf{A}$ matrices.
\end{theorem}

\begin{IEEEproof}
	The proof is given in Appendix~\ref{app:theorem_2}.
\end{IEEEproof}

Theorem~\ref{thm:FCFP} essentially says that we can replace the standard multi-dimensional ratios in~\eqref{q} by their quadratic transformed terms as in Lemma~\ref{lem:quadratic_transform}, provided that the outer functions $\mathsf{h}^{\mathsf{o}}(\cdot)$ and $\mathsf{h}^{\mathsf{c}}_k(\cdot)$ are nondecreasing on their respective domains. The new problem~\eqref{equivalent Ray} has additional constraints as compared to the original problem, i.e., constraints~\eqref{range_01} and~\eqref{range_02}. Note that any $\mathbf{x} \in \mathcal X$ together with their corresponding optimal $\boldsymbol{\lambda}^{\mathsf{o}}_i$ and $\boldsymbol{\lambda}^{\mathsf{c}}_{k,j}$ are contained in the new constraints, because of \eqref{condition_01} and \eqref{condition_02}.
These constraints are needed to ensure that the values of the transformed terms are within the domains of the outer functions where the functions are nondecreasing in each component. 

When applying Theorem~\ref{thm:FCFP} to reformulate FCFP problems as given in~\eqref{q}, we must specify the appropriate domains for the outer functions to ensure that the conditions of the theorem are satisfied.
We remark that when the outer functions are nondecreasing over the entire space, we can choose $\mathcal{H}^{\mathsf{o}} = \mathbb{R}^{I}$ and $\mathcal{H}_k^{\mathsf{c}} = \mathbb{R}^{J}$. 
In this case, the constraints in \eqref{range_01} and \eqref{range_02} can be omitted.

We now apply Theorem~\ref{thm:FCFP} to transform problem~(P1).
The objective function $\mathcal{A}(\mathbf{x})$ is a weighted sum of standard multi-dimensional ratios.
Since all the weights are positive, the outer function, i.e., the weighted summation, is nondecreasing over the entire space, so $\mathcal{H}^{\mathsf{o}} = \mathbb{R}^{R}$.
Moreover, since the constraint functions are standard multi-dimensional ratios and not composed with any outer functions, so $\mathcal{H}_k^{\mathsf{c}} = \mathbb{R}$.
Then, the constraints \eqref{range_01} and \eqref{range_02} can be omitted when applying Theorem~\ref{thm:FCFP} to (P1), yielding
\begin{subequations} \label{28}
	\begin{align}
		\textbf{{(P2):}} \;\;\;
		\mathop{\mathrm{maximize}}_{ \mathbf{x}, \boldsymbol{\lambda}_r^{\mathsf{o}},  \boldsymbol{\lambda}_k^{\mathsf{c}}} \;\;
		&\;
		\tilde{\mathcal{A}} (\mathbf{x}, \boldsymbol{\lambda}_r^{\mathsf{o}})
		\triangleq
		\sum_{r=1}^{R} \kappa_r \tilde{\mathsf{f}}_r^{\mathsf{o}} (\mathbf{x}, \boldsymbol{\lambda}_r^{\mathsf{o}})  \label{newObj}  \\
		\mathop{\mathrm{subject\;to}} \;\;
		& \;
		\tilde{\mathsf{f}}_k^{\mathsf{c}} (\mathbf{x}, \boldsymbol{\lambda}_k^{\mathsf{c}}) \geq \Gamma_k, \ \forall k,  \\
		& 
		\left| x_n \right| = 1, \ \forall n,  \label{CMs}
	\end{align}
\end{subequations}
where the transformed functions are given by
\begin{align}
	\tilde{\mathsf{f}}_r^{\mathsf{o}} (\mathbf{x}, \boldsymbol{\lambda}_r^{\mathsf{o}}) 
	& = 
	2 \mathop{\mathfrak{Re}} \! \left\{ (\boldsymbol{\lambda}_{r}^{\mathsf{o}})^{\dagger} \mathbf{R}_{r} \mathbf{x} \right\} 
	- (\boldsymbol{\lambda}_{r}^{\mathsf{o}})^{\dagger} \mathbf{\Sigma}^{\mathsf{o}}(\mathbf{x}) \boldsymbol{\lambda}_r^{\mathsf{o}},  \label{30}  \\
	\tilde{\mathsf{f}}_k^{\mathsf{c}} (\mathbf{x}, \boldsymbol{\lambda}_{k}^{\mathsf{c}}) 
	& = 
	2 \mathop{\mathfrak{Re}} \! \left\{ (\boldsymbol{\lambda}_{k}^{\mathsf{c}})^{\dagger} \mathbf{H}_{k} \mathbf{x} \right\} 
	- (\boldsymbol{\lambda}_{k}^{\mathsf{c}})^{\dagger} \mathbf{\Sigma}^{\mathsf{c}}_{k}(\mathbf{x}) \boldsymbol{\lambda}_{k}^{\mathsf{c}}.  \label{sinrFP}
\end{align}
Problem (P2) is convex in $\mathbf{x}$ except for the extra constant-modulus constraints \eqref{CMs}.
To verify convexity, we first examine the function (\ref{30}).
The first term in \eqref{30} is affine in $\mathbf{x}$;
the second term in \eqref{30} can be rewritten as
\begin{equation}
	(\boldsymbol{\lambda}_{r}^{\mathsf{o}})^{\dagger} \mathbf{\Sigma}^{\mathsf{o}}(\mathbf{x}) \boldsymbol{\lambda}_{r}^{\mathsf{o}}
	= 
	\mathbf{x}^{\dagger} \mathbf{M}_{r}^{\mathsf{o}}(\boldsymbol{\lambda}_{r}^{\mathsf{o}}) \mathbf{x} 
	+ 
	\sigma_n^2 \left\| \boldsymbol{\lambda}_{r}^{\mathsf{o}} \right\|_2^2,
\end{equation}
where the coefficient of the quadratic term is
\begin{align} \label{Psi}
	\mathbf{M}_{r}^{\mathsf{o}} (\boldsymbol{\lambda}_{r}^{\mathsf{o}})
	& = 
	\sum_{k = 1}^{K} p_{k}
	\left( \mathbf{H}_{k}^{\dagger} \boldsymbol{\lambda}_{r}^{\mathsf{o}} \right)
	\left( \mathbf{H}_{k}^{\dagger} \boldsymbol{\lambda}_{r}^{\mathsf{o}} \right)^{\dagger}.
\end{align}
Clearly, $\mathbf{M}_{r}^{\mathsf{o}}(\boldsymbol{\lambda}_{r}^{\mathsf{o}})$ is positive semidefinite.
Hence, the function \eqref{30} is concave with respect to $\mathbf{x}$, as
\begin{multline}
	\tilde{\mathsf{f}}_r^{\mathsf{o}} (\mathbf{x}, \boldsymbol{\lambda}_{r}^{\mathsf{o}}) = \\
	- \mathbf{x}^{\dagger} \mathbf{M}_{r}^{\mathsf{o}}(\boldsymbol{\lambda}_{r}^{\mathsf{o}}) \mathbf{x}
	+ 2 \mathop{\mathfrak{Re}} \! \left\{ (\boldsymbol{\lambda}_{r}^{\mathsf{o}})^{\dagger} \mathbf{R}_r \mathbf{x} \right\}
	- \sigma_n^2 \left\| \boldsymbol{\lambda}_{r}^{\mathsf{o}} \right\|_2^2.
\end{multline}
Similarly, \eqref{sinrFP} can be rewritten as
\begin{multline}
	\tilde{\mathsf{f}}_k^{\mathsf{c}} (\mathbf{x}, \boldsymbol{\lambda}_{k}^{\mathsf{c}}) = \\
	- 
	\mathbf{x}^{\dagger} \mathbf{M}_{k}^{\mathsf{c}}(\boldsymbol{\lambda}_{k}^{\mathsf{c}}) \mathbf{x}
	+ 
	2 \mathop{\mathfrak{Re}} \! \left\{ (\boldsymbol{\lambda}_{k}^{\mathsf{c}})^{\dagger} \mathbf{H}_{k} \mathbf{x} \right\}
	- \sigma_n^2 \left\| \boldsymbol{\lambda}_{k}^{\mathsf{c}} \right\|_2^2,
\end{multline}
where $\mathbf{M}_{k}^{\mathsf{c}}(\boldsymbol{\lambda}_{k}^{\mathsf{c}})$ is positive semidefinite and is given by
\begin{multline} \label{Lambda}
	\mathbf{M}_{k}^{\mathsf{c}} (\boldsymbol{\lambda}_{k}^{\mathsf{c}})
	= 
	\sum_{j \neq k} p_{j}
	\left( \mathbf{H}_{j}^{\dagger} \boldsymbol{\lambda}_{k}^{\mathsf{c}} \right)
	\left( \mathbf{H}_{j}^{\dagger} \boldsymbol{\lambda}_{k}^{\mathsf{c}} \right)^{\dagger}  \\
	+ p
	\mathop{\mathbb{E}_{q \left( \eta \right)}} \!
	\left[ \left( \alpha \mathbf{U}^{\dagger} ( \eta ) \boldsymbol{\lambda}_{k}^{\mathsf{c}} \right) \left( \alpha \mathbf{U}^{\dagger} ( \eta ) \boldsymbol{\lambda}_{k}^{\mathsf{c}} \right)^{\dagger} \right].
\end{multline}
The expectation in \eqref{Lambda} can be rewritten in a manner similar to \eqref{reWr}, so that the auxiliary variable $\boldsymbol{\lambda}_{k}^{\mathsf{c}}$ can be extracted from the expectation operation.
In the next subsection, we handle the constant-modulus constraints in problem (P2).

\subsection{Penalty-Based Method for Modulus Constraints} \label{Section IV-B}

To tackle the constant-modulus constraints \eqref{CMs}, we adopt a penalty-based approach as a heuristic solution.
Specifically, we relax the constraint \eqref{CMs} to $\left| x_n \right| \leq 1$, $\forall n$, then introduce a penalty term into the objective to encourage the solution to be constant-modulus.
By doing so, (P2) is transformed into
\begin{subequations} \label{49}
	\begin{align}
		\mathop{\mathrm{maximize}}_{\mathbf{x}, \mathbf{z}, \boldsymbol{\lambda}_r^{\mathsf{o}}, \boldsymbol{\lambda}_k^{\mathsf{c}}} \;\;
		&\;
		\tilde{\mathcal{A}} (\mathbf{x}, \boldsymbol{\lambda}_r^{\mathsf{o}})
		- \mu \left\| \mathbf{x} - \mathbf{z} \right\|_2^2  \label{49a}  \\
		\mathop{\mathrm{subject\;to}} \;\;
		&\;
		\tilde{\mathsf{f}}_k^{\mathsf{c}} (\mathbf{x}, \boldsymbol{\lambda}_k^{\mathsf{c}}) \geq \Gamma_k, \ \forall k,  \label{49b}  \\
		& \left| x_n \right| \leq 1, \ \left| z_n \right| = 1, \ \forall n,
	\end{align}
\end{subequations}
where $\mu > 0$ represents the coefficient of the penalty term and the new auxiliary variable $\mathbf{z}$ is unit-modulus.
This problem can be solved by iteratively updating the RIS reflecting coefficients and all the other auxiliary variables, as shown in Algorithm~\ref{algorithm_1}.
Note that if the value of $\mu$ is too small, the penalty term may be insufficient to enforce the solution to satisfy the constant-modulus constraint, whereas a large $\mu$ enforces the constraint more strictly but significantly slows down convergence.
Hence, we first initialize $\mu$ to a small positive value and then gradually increase it in an outer loop until the penalty term is zero.

\begin{algorithm}[!t]
	
	\caption{\strut Quadratic Transform and Penalty Based Method}
	
	\begin{algorithmic}[1] \label{algorithm_1}
		\vspace{3pt}
		
		\STATE Initialize $\mu > 0$.
		\STATE Initialize $\mathrm{num\_iter} = 0$ and $\mathrm{max\_iter}$. 
		\STATE Initialize $\mathbf{x}$ as a feasible point of the original problem (P1).
		
		\REPEAT
		
		\REPEAT
		
		\STATE Update $\mathbf{z} = \exp \left[ j \mathop{\mathrm{arg}} \left( \mathbf{x} \right) \right]$.
		
		\STATE Update $\boldsymbol{\lambda}_{r}^{\mathsf{o}} = (\mathbf{\Sigma}^{\mathsf{o}} (\mathbf{x}))^{-1} \mathbf{R}_r \mathbf{x}$ for all $r$.
		
		\STATE Update $\boldsymbol{\lambda}_{k}^{\mathsf{c}} = (\mathbf{\Sigma}^{\mathsf{c}}_{k}(\mathbf{x}))^{-1} \mathbf{H}_{k} \mathbf{x}$ for all $k$.
		
		\STATE Optimize $\mathbf{x}$ in \eqref{49} with fixed auxiliary variables.
		
		\UNTIL{convergence} 
		
		\STATE $\mu = \xi \mu$ with $\xi > 1$.
		
		\STATE $\mathrm{num\_iter} = \mathrm{num\_iter} + 1$.
		
		\UNTIL{$\left\| \mathbf{x} - \mathbf{z} \right\| = 0$ or $\mathrm{num\_iter} > \mathrm{max\_iter}$}.
		
		\STATE Update $\mathbf{x} = \exp \left[j \mathop{\mathrm{arg}} \left( \mathbf{x} \right) \right]$.
		
		\STATE \textbf{output} The optimized RIS reflection coefficients $\mathbf{x}$.
		
	\end{algorithmic}
	
\end{algorithm}

To show the convergence of the proposed algorithm, we first consider the inner loop, i.e., the algorithm with a fixed $\mu$.
The key is that in each iteration of the inner loop, all the auxiliary variables are updated optimally so that both the values of the objective and the constraint functions are \emph{nondecreasing}.
Thus, the solution of $\mathbf{x}$ updated in the previous iteration is \emph{feasible} in the current iteration.
Furthermore, optimizing $\mathbf{x}$ in \eqref{49} with fixed auxiliary variables is a \emph{convex} quadratically constrained quadratic program (QCQP), the optimal solution of which can be easily obtained by an optimization solver, such as the CVX.
Hence, the update of $\mathbf{x}$ in the current iteration leads to a non\-decreasing objective value, so the inner loop must converge.

In the outer loop, the penalty increases as $\mu$ increases, making the solution more likely to satisfy the modulus constraints.
If a penalty coefficient $\mu$ is selected so that the penalty term is zero when the algorithm converges, the Karush-Kuhn-Tucker (KKT) conditions on $\mathbf{x}$ for the problem \eqref{49}, under the optimal auxiliary variables, must be the same as for the problem (P1), hence the algorithm must have converged to a KKT point of the original problem (P1).

The computational complexity of the proposed algorithm in each iteration mainly arises from solving a QCQP problem of updating $\mathbf{x}$, thus the per-iteration computational complexity of this algorithm is dominated by $\mathcal{O} \left( N^3 + K N^2 \right)$.

\section{Constant-Modulus Linear Transform Based Algorithm for RIS Reflection Coefficient Design}
\label{Section_05}

In the previous section, we employ the quadratic transform to deal with the fractional structures and utilize a penalty-based method to handle constant-modulus constraints.
However, the computational complexity of solving a QCQP problem in each iteration is still substantial.
Furthermore, to determine a proper penalty coefficient, an extra outer loop is required, which also increases the computational complexity.
To reduce complexity, a new transform, called the constant-modulus linear transform, is proposed in this section to solve problem (P1), building upon the quadratic transform and the constant-modulus property.

\subsection{Constant-Modulus Linear Transform}

The following lemma is the basis of the new transform.

\begin{lemma} \label{lem:linear_transform}
	A lower bound for a standard multi-dimensional ratio function $\mathsf{f}(\mathbf{x})$ in \eqref{ratio} with variable $\mathbf{x}$ being \emph{unit-modulus} can be constructed as 
	\begin{align} \label{transL} \hspace{-5pt}
		\mathsf{f} (\mathbf{x})
		& \geq
		\bar{\mathsf{f}} (\mathbf{x}, \mathbf{z}, \boldsymbol{\lambda})  \notag  \\
		& \triangleq
		2 \mathop{\mathfrak{Re}} \! 
		\left\lbrace
			\mathbf{x}^{\dagger} 
			\left(
				\left( \delta \mathbf{I} - \mathbf{M}(\boldsymbol{\lambda}) \right) \mathbf{z} 
				+ 
				\mathbf{A}^{\dagger} \boldsymbol{\lambda} 
			\right)
		\right\rbrace
		+ 
		c \left( \mathbf{z}, \boldsymbol{\lambda} \right),
	\end{align}
	for all $\boldsymbol{\lambda} \in \mathbb{C}^{M}$ and $\mathbf{x}, \mathbf{z} \in \mathbb{U}^{N}$,
	where $\mathbb{U} \triangleq \left\lbrace x \in \mathbb{C} \mathop{|} |x| = 1 \right\rbrace$,
	the matrix $\mathbf{M}(\boldsymbol{\lambda})$ is given by
	\begin{equation}
		\mathbf{M} (\boldsymbol{\lambda})
		= 
		\sum_{m} 
		\varrho_m
		\left( \mathbf{B}_{m}^{\dagger} \boldsymbol{\lambda} \right) 
		\left( \mathbf{B}_{m}^{\dagger} \boldsymbol{\lambda} \right)^{\dagger},
	\end{equation}
	the parameter $\delta$ is the trace of $\mathbf{M} (\boldsymbol{\lambda})$, 
	and $c \left( \mathbf{z}, \boldsymbol{\lambda} \right)$ is given by
	\begin{equation} \label{constant_c}
		c \left( \mathbf{z}, \boldsymbol{\lambda} \right)
		=
		\mathbf{z}^{\dagger} \mathbf{M}(\boldsymbol{\lambda}) \mathbf{z} 
		- 
		2 \delta N 
		-  
		\boldsymbol{\lambda}^{\dagger} \mathbf{C} \boldsymbol{\lambda} .
	\end{equation}
	The equality in (\ref{transL}) is achieved at $\mathbf{z}^{\star} = \mathbf{x}$ and $\boldsymbol{\lambda}^{\star}$ in (\ref{op_aux1}).
\end{lemma}
\begin{IEEEproof}
	The proof is given in Appendix~\ref{app:lemma_2}.
\end{IEEEproof}
This lemma uses a key technique in \cite{9931490} that turns a  quadratic optimization over $\mathbf{x}$ with unit-modulus constraints into a linear optimization. 
This is made possible by taking advantage of the unit-modulus property of the variable $\mathbf{x}$.

Based on Lemma~\ref{lem:linear_transform}, the constant-modulus linear transform, which is proposed for FCFP problems with constant-modulus constraints, can be stated in the following theorem.

\begin{theorem} \label{thm:linear_transform}
	Consider a maximization FCFP problem with constant-modulus constraints
	\begin{subequations} \label{FCFP_unit}
		\begin{align}
			\mathop{\mathrm{maximize}}_{\mathbf{x}} \;\;
			&
			\sum_{i} w_i \mathsf{f}_i^{\mathsf{o}} (\mathbf{x}) \\
			\mathop{\mathrm{subject\;to}} \;\;
			& \;
			\mathsf{f}_j^{\mathsf{c}} (\mathbf{x}) \geq c_j, \ \forall j, \\
			&
			\left| x_n \right| = 1, \ \forall n,
		\end{align}
	\end{subequations}
	where $w_i > 0$ denotes the weight, $\mathsf{f}^{\mathsf{o}}_i (\mathbf{x})$ and $\mathsf{f}^{\mathsf{c}}_j(\mathbf{x})$ are multi-dimensional ratio functions of the form in \eqref{ratio}. 
	This problem is equivalent to
	\begin{subequations} \label{newLP}
		\begin{align}
			\mathop{\mathrm{maximize}}_{\mathbf{x}, \mathbf{z}, \boldsymbol{\lambda}_i^{\mathsf{o}}, \boldsymbol{\lambda}_j^{\mathsf{c}}} \;\;
			&
			\sum_{i} 
			w_i \bar{\mathsf{f}}_i^{\mathsf{o}} (\mathbf{x}, \mathbf{z}, \boldsymbol{\lambda}^{\mathsf{o}}_i)  \\
			\mathop{\mathrm{subject\;to}} \;\;
			& \;
			\bar{\mathsf{f}}_j^{\mathsf{c}} (\mathbf{x}, \mathbf{z}, \boldsymbol{\lambda}^{\mathsf{c}}_j) \geq c_j, \ \forall j, \label{48b} \\
			&
			\left| x_n \right| = \left| z_n \right| = 1, \ \forall n,
		\end{align}
	\end{subequations}
	where the transformed functions $\bar{\mathsf{f}}_i^{\mathsf{o}} (\mathbf{x}, \mathbf{z}, \boldsymbol{\lambda}^{\mathsf{o}}_i)$ and $\bar{\mathsf{f}}_j^{\mathsf{c}} (\mathbf{x}, \mathbf{z}, \boldsymbol{\lambda}^{\mathsf{o}}_j)$ are defined as in \eqref{transL}.
	The optimal $\mathbf{z}$ equals to the optimal $\mathbf{x}$, and the optimal $\boldsymbol{\lambda}^{\mathsf{o}}_i$ and $\boldsymbol{\lambda}^{\mathsf{c}}_j$ are functions of the optimal $\mathbf{x}$ in the form of \eqref{op_aux1} with their respective $\mathbf{D}(\mathbf{x})$ and $\mathbf{A}$ matrices.
\end{theorem}

\begin{IEEEproof}
	Based on Lemma~\ref{lem:linear_transform}, all the ratio functions in both the objective and the constraints are maximized with the same value for the variable $\mathbf{z}$, i.e., $\mathbf{z}^{\star} = \mathbf{x}$.
	This must be the optimal solution of $\mathbf{z}$ to problem \eqref{newLP}, since it not only maximizes the objective function but also yields \emph{the largest constraint set} for the variable $\mathbf{x}$.
	Substituting $\mathbf{z}^{\star} = \mathbf{x}$ into \eqref{newLP} yields
	\begin{subequations}
		\begin{align}
			\mathop{\mathrm{maximize}}_{\mathbf{x}, \boldsymbol{\lambda}_i^{\mathsf{o}}, \boldsymbol{\lambda}_j^{\mathsf{c}}} \;\;
			&
			\sum_{i} 
			w_i \bar{\mathsf{f}}_i^{\mathsf{o}} (\mathbf{x}, \mathbf{z}^{\star}, \boldsymbol{\lambda}^{\mathsf{o}}_i)
			=
			\sum_{i} 
			w_i \tilde{\mathsf{f}}_i^{\mathsf{o}} (\mathbf{x}, \boldsymbol{\lambda}^{\mathsf{o}}_i) \\
			\mathop{\mathrm{subject\;to}} \;\;
			& \;
			\bar{\mathsf{f}}_j^{\mathsf{c}} (\mathbf{x}, \mathbf{z}^{\star}, \boldsymbol{\lambda}^{\mathsf{c}}_j)
			=
			\tilde{\mathsf{f}}_j^{\mathsf{c}} (\mathbf{x}, \boldsymbol{\lambda}^{\mathsf{c}}_j) \geq c_j, \ \forall j, \\
			&
			\left| x_n \right| = 1, \ \forall n,
		\end{align}
	\end{subequations}
	where the functions $\tilde{\mathsf{f}}^{\mathsf{o}}_i (\mathbf{x}, \boldsymbol{\lambda}^{\mathsf{o}}_i)$, $\tilde{\mathsf{f}}^{\mathsf{c}}_{j} (\mathbf{x}, \boldsymbol{\lambda}^{\mathsf{c}}_{j})$  are defined as in \eqref{Minor1}.
	The rest of the proof follows by invoking Theorem~\ref{thm:FCFP}.
\end{IEEEproof}

\subsection{Reformulating (P1) via the Proposed Transform}

Based on Theorem~\ref{thm:linear_transform}, problem (P1) is equivalent to
\begin{subequations} \label{70}
	\begin{align}
		\textbf{(P3):} \;\;\;
		\mathop{\mathrm{maximize}}_{\mathbf{x}, \mathbf{z}, \boldsymbol{\lambda}_r^{\mathsf{o}}, \boldsymbol{\lambda}_k^{\mathsf{c}}} \;\;
		& \;
		\bar{\mathcal{A}} (\mathbf{x}, \mathbf{z}, \boldsymbol{\lambda}_r^{\mathsf{o}})
		\triangleq
		\sum_{r=1}^{R} 
		\kappa_r 
		\bar{\mathsf{f}}_r^{\mathsf{o}} (\mathbf{x}, \mathbf{z}, \boldsymbol{\lambda}_r^{\mathsf{o}})  \\
		\mathop{\mathrm{subject\;to}} \;\;
		& \;
		\bar{\mathsf{f}}_k^{\mathsf{c}} (\mathbf{x}, \mathbf{z}, \boldsymbol{\lambda}_k^{\mathsf{c}}) \geq \Gamma_k, \ \forall k, \\
		&
		\left| x_n \right| = \left| z_n \right| = 1, \ \forall n.
	\end{align}
\end{subequations}
The transformed functions in problem (P3) are given by
\begin{align}
	\bar{\mathsf{f}}_r^{\mathsf{o}} & (\mathbf{x}, \mathbf{z}, \boldsymbol{\lambda}_r^{\mathsf{o}}) = \notag \\
	& 2 \mathop{\mathfrak{Re}} \! 
	\Big\lbrace
		\mathbf{x}^{\dagger} 
		\Big(\!
			\left( \delta_r^{\mathsf{o}} \mathbf{I} - \mathbf{M}_{r}^{\mathsf{o}} (\boldsymbol{\lambda}_{r}^{\mathsf{o}}) \right) \mathbf{z} + \mathbf{R}_r^{\dagger} \boldsymbol{\lambda}_r^{\mathsf{o}} 
		\Big)
	\Big\rbrace 
	+ {c}^{\mathsf{o}}_r \left( \mathbf{z}, \boldsymbol{\lambda}_r^{\mathsf{o}} \right),
\end{align}
\begin{align}
	\bar{\mathsf{f}}_k^{\mathsf{c}} & (\mathbf{x}, \mathbf{z}, \boldsymbol{\lambda}_k^{\mathsf{c}}) = \notag \\
	& 2 \mathop{\mathfrak{Re}} \! 
	\Big\lbrace
		\mathbf{x}^{\dagger} 
		\Big(\!
			\left( \delta_{k}^{\mathsf{c}} \mathbf{I} - \mathbf{M}_{k}^{\mathsf{c}} (\boldsymbol{\lambda}_{k}^{\mathsf{c}}) \right) \mathbf{z} + \mathbf{H}_{k}^{\dagger} \boldsymbol{\lambda}_{k}^{\mathsf{c}} 
		\Big)
	\Big\rbrace 
	+ 
	{c}^{\mathsf{c}}_k \left( \mathbf{z}, \boldsymbol{\lambda}_{k}^{\mathsf{c}} \right) ,  
\end{align}
where the matrices $\mathbf{M}_{r}^{\mathsf{o}}(\boldsymbol{\lambda}_{r}^{\mathsf{o}})$ and $\mathbf{M}_{k}^{\mathsf{c}}(\boldsymbol{\lambda}_{k}^{\mathsf{c}})$ are expressed in \eqref{Psi} and \eqref{Lambda}, respectively, 
$\delta_r^{\mathsf{o}}$ and $\delta_k^{\mathsf{c}}$ represent the traces of the matrices $\mathbf{M}_{r}^{\mathsf{o}}(\boldsymbol{\lambda}_{r}^{\mathsf{o}})$ and $\mathbf{M}_{k}^{\mathsf{c}}(\boldsymbol{\lambda}_{k}^{\mathsf{c}})$, respectively,
and ${c}^{\mathsf{o}}_r \left( \mathbf{z}, \boldsymbol{\lambda}_r^{\mathsf{o}} \right)$ and $c^{\mathsf{c}}_k \left( \mathbf{z}, \boldsymbol{\lambda}_{k}^{\mathsf{c}} \right)$ are given by
\begin{align}
	c^{\mathsf{o}}_r \left( \mathbf{z}, \boldsymbol{\lambda}_r^{\mathsf{o}} \right) 
	& = 
	\mathbf{z}^{\dagger} \mathbf{M}_{r}^{\mathsf{o}}(\boldsymbol{\lambda}_{r}^{\mathsf{o}}) \mathbf{z} 
	- 
	2 \delta_{r}^{\mathsf{o}} N
	-  
	\sigma_n^2 \left\| \boldsymbol{\lambda}_r^{\mathsf{o}} \right\|_2^2, \\
	c^{\mathsf{c}}_k \left( \mathbf{z}, \boldsymbol{\lambda}_{k}^{\mathsf{c}} \right)
	& = 
	\mathbf{z}^{\dagger} \mathbf{M}_{k}^{\mathsf{c}}(\boldsymbol{\lambda}_{k}^{\mathsf{c}}) \mathbf{z}
	- 
	2 \delta_{k}^{\mathsf{c}} N
	- 
	\sigma_n^2 \left\| \boldsymbol{\lambda}_k^{\mathsf{c}} \right\|_2^2.
\end{align}
Similar to the problem \eqref{49}, problem (P3) can be tackled via iteratively updating the RIS reflection coefficients $\mathbf{x}$ and the auxiliary variables.
When all the auxiliary variables are fixed, updating $\mathbf{x}$ is an LP with constant-modulus constraints as
\begin{subequations}
	\begin{align}
		\textbf{(P4):} \;\;\;
		\mathop{\mathrm{maximize}}_{\mathbf{x}} \;\;
		&
		\sum_{r=1}^{R} 
		\kappa_r \bar{\mathsf{f}}_r^{\mathsf{o}} (\mathbf{x}, \mathbf{z}, \boldsymbol{\lambda}_r^{\mathsf{o}}) \label{55a} \\
		\mathop{\mathrm{subject\;to}} \;\;
		& \;
		\bar{\mathsf{f}}_k^{\mathsf{c}} (\mathbf{x}, \mathbf{z}, \boldsymbol{\lambda}_k^{\mathsf{c}}) \geq \Gamma_k, \ \forall k, \label{55b} \\
		&
		\left| x_n \right| = 1 , \ \forall n. \label{55c}
	\end{align}
\end{subequations}
When $\mathbf{x}$ is held fixed, all the optimal auxiliary variables can be found in closed forms.
The details are shown in Algorithm~\ref{algorithm_2}.

\begin{algorithm}[t]
	\caption{\strut Linear Transform Based Method}
	\begin{algorithmic}[1] \label{algorithm_2}
		\vspace{3pt}
		
		\STATE Initialize $\mathbf{x}$ as a feasible point of the original problem (P1).
		
		\REPEAT
		
		\STATE Update $\mathbf{z} = \mathbf{x}$.
		
		\STATE Update $\boldsymbol{\lambda}_{r}^{\mathsf{o}} = (\mathbf{\Sigma}^{\mathsf{o}} (\mathbf{x}))^{-1} \mathbf{R}_r \mathbf{x}$ for all $r$. 
		
		\STATE Update $\boldsymbol{\lambda}_{k}^{\mathsf{c}} = (\mathbf{\Sigma}^{\mathsf{c}}_{k}(\mathbf{x}))^{-1} \mathbf{H}_{k} \mathbf{x}$ for all $k$.
		
		\STATE Update $\mathbf{x}$ by optimally solving problem (P4).
		
		\UNTIL{convergence}
		
		\STATE \textbf{output} The optimized RIS reflection coefficients $\mathbf{x}$.
	\end{algorithmic}
\end{algorithm}

Note that the convergence of the proposed linear transform based algorithm, i.e., Algorithm~\ref{algorithm_2}, can be analyzed in a similar manner as that of Algorithm~\ref{algorithm_1}, and it is therefore omitted here.
The convergence of Algorithm~\ref{algorithm_2} to a KKT solution of problem (P1) can be guaranteed if problem (P4) can be solved globally optimally.
In the sequel, we show that the global optimality of problem (P4) can be achieved with low complexity scaling in the number of RIS reflecting elements, provided that a certain condition is satisfied.

\subsection{Achieving Global Optimality of (P4)}

The work \cite{9931490} establishes strong duality for an LP with constant-modulus constraints under certain conditions.
In this section, we provide a similar but simpler way of obtaining the globally optimal solution to problem (P4).
Consider first a relaxed version of problem (P4) obtained by replacing the non-convex constraint \eqref{55c} with the constraint $\left| x_n \right| \leq 1$, $\forall n$,\begin{subequations} \label{relaxedP}
	\begin{align}
		\mathop{\mathrm{maximize}}_{\mathbf{x}} \;\;
		&
		\sum_{r=1}^{R} 
		\kappa_r \bar{\mathsf{f}}_r^{\mathsf{o}} (\mathbf{x}, \mathbf{z}, \boldsymbol{\lambda}_r^{\mathsf{o}}) \\
		\mathop{\mathrm{subject\;to}} \;\;
		&\;
		\bar{\mathsf{f}}_k^{\mathsf{c}} (\mathbf{x}, \mathbf{z}, \boldsymbol{\lambda}_k^{\mathsf{c}}) \geq \Gamma_k, \ \forall k, \\
		&
		\left| x_n \right| \leq 1, \ \forall n.
	\end{align}
\end{subequations} 
Note that the relaxed problem \eqref{relaxedP} is convex and enjoys strong duality. 
Its dual problem is given by
\begin{align} \label{dual}
	\mathop{\mathrm{minimize}}_{\boldsymbol{\nu} \geq\mathop{} \mathbf{0}} \; \mathop{\mathrm{max}}_{ |x_n| \leq 1, \forall n } \;
	\mathfrak{L} \left( \mathbf{x}, \boldsymbol{\nu} \right)
\end{align}
where $\boldsymbol{\nu} = [\nu_1, \ldots, \nu_K]^{\mathsf{T}}$ denotes the dual variable associated with the linear constraints, and the partial Lagrangian is
\begin{equation} \label{Lagrange}
	\mathfrak{L} \left( \mathbf{x}, \boldsymbol{\nu} \right)
	= 
	\sum_{r=1}^{R} 
	\kappa_r \bar{\mathsf{f}}_r^{\mathsf{o}} (\mathbf{x}, \mathbf{z}, \boldsymbol{\lambda}_r^{\mathsf{o}})  
	+
	\sum_{k=1}^{K} 
	\nu_k \left( \bar{\mathsf{f}}_k^{\mathsf{c}} (\mathbf{x}, \mathbf{z}, \boldsymbol{\lambda}_k^{\mathsf{c}}) - \Gamma_k \right).
\end{equation}
Since \eqref{Lagrange} is linear over $\mathbf{x}$, for fixed $\boldsymbol{\nu}$, the inner maximization in \eqref{dual} is achieved if the entries of $\mathbf{x}$ can be set to match the phases of the overall linear coefficient $\boldsymbol{\varsigma} (\boldsymbol{\nu})$, i.e.,
\begin{align} \label{opt_x}
	\mathbf{x} (\boldsymbol{\nu})
	& = 
	\exp \left( j \arg \left( \sum_{r=1}^{R} \kappa_r \mathbf{d}_{r}^{\mathsf{o}} + \sum_{k=1}^{K} \nu_k \mathbf{d}_{k}^{\mathsf{c}} \right) \right)  \notag \\
	& \triangleq 
	\exp \left( j \arg \left( \boldsymbol{\varsigma} (\boldsymbol{\nu}) \right) \right),
\end{align}
where $\mathbf{d}_{r}^{\mathsf{o}}$ and $\mathbf{d}_{k}^{\mathsf{c}}$ are the liner coefficients in $\bar{\mathsf{f}}_r^{\mathsf{o}} (\mathbf{x}, \mathbf{z}, \boldsymbol{\lambda}_r^{\mathsf{o}})$ and $\bar{\mathsf{f}}_k^{\mathsf{c}} (\mathbf{x}, \mathbf{z}, \boldsymbol{\lambda}_k^{\mathsf{c}})$, respectively.
In this case, the constraint $|x_n| \le 1$ is achieved with equality.
The dual for the relaxed problem attains the same optimal value as the problem below 
\begin{align} \label{60}
	\mathop{\mathrm{minimize}}_{\boldsymbol{\nu} \geq \mathbf{0}} \; \mathop{\mathrm{max}}_{|x_n| = 1, \forall n} \;
	\mathfrak{L} \left( \mathbf{x}, \boldsymbol{\nu} \right),
\end{align}
which is the dual problem of the original problem (P4). 
Since the optimal value of the dual problem is an upper bound to the optimal value of (P4), as long as this upper bound is achieved with $x_n$ that satisfies the constraint $|x_n|=1$, such $x_n$ must be the global optimal solution to (P4).

It remains to check whether we can find a primal solution $\mathbf{x}(\boldsymbol{\nu})$ for problem (P4) that satisfies $|x_n|=1$.
Such a solution is given in \eqref{opt_x}, but it can be readily found if
\begin{align} \label{mild_condition}
	\left[ \boldsymbol{\varsigma} ( \boldsymbol{\nu}^{\star} ) \right]_n \neq 0, \ \forall n,
\end{align}
where $\boldsymbol{\nu}^{\star}$ denotes the optimal dual solution.
Thus, if the above condition is satisfied, we get a global optimal solution to (P4).
We summarize this statement below.

\begin{proposition}
	If the optimal solution $\boldsymbol{\nu}^{\star}$ to the dual problem of problem (P4) satisfies the condition \eqref{mild_condition}, then the solution given by \eqref{opt_x} is the globally optimal solution to problem (P4).
\end{proposition}

The above proposition potentially provides an efficient method to obtain the globally optimal solution to (P4), even though (P4) is non-convex.
Specifically, substituting $\mathbf{x} (\boldsymbol{\nu})$ in \eqref{opt_x} into the dual objective function in \eqref{60}, we obtain
\begin{align}
	\mathfrak{L} \left( \mathbf{x} (\boldsymbol{\nu}), \boldsymbol{\nu} \right)
	= 
	2 & \left\| \sum_{r=1}^{R} \kappa_r \mathbf{d}_{r}^{\mathsf{o}} + \sum_{k=1}^{K} \nu_k \mathbf{d}_{k}^{\mathsf{c}} \right\|_{1} 
	+
	\sum_{r=1}^{R} \kappa_r {c}^{\mathsf{o}}_r \left( \mathbf{z},  \boldsymbol{\lambda}_r^{\mathsf{o}} \right) \notag \\
	& + 
	\sum_{k=1}^{K} \nu_k \left( {c}^{\mathsf{c}}_k \left( \mathbf{z}, \boldsymbol{\lambda}_{k}^{\mathsf{c}} \right) - \Gamma_k \right),
\end{align}
which is convex in $\boldsymbol{\nu}$.
In fact, the dual problem \eqref{60} is an LP, so it can be easily solved.
For example, it can be solved using the coordinate bisection algorithm \cite{9931490}, the alternating direction method of multipliers (ADMM) \cite{boyd2011admm}, or directly by an efficient optimization solver such as MOSEK \cite{mosek2019}.
In the next step, we verify whether the obtained solution $\boldsymbol{\nu}^{\star}$ satisfies the condition \eqref{mild_condition}. 
If the condition \eqref{mild_condition} is satisfied, the globally optimal solution of problem (P4) can be directly obtained from the closed-form solution \eqref{opt_x}.

Empirically, in the numerical simulations of the scenarios considered in this paper, where the number of RIS reflecting elements is much greater than the number of users, the condition \eqref{mild_condition} is almost always satisfied.
A similar observation is made in \cite[Appendix D]{9931490}.

The per-iteration computational complexity of the constant-modulus linear transform based algorithm is dominated by that of solving the LP \eqref{60}, which is of size $\mathcal{O} \left( K \right)$, but the evaluation of its objective takes $N$ operations.
Specifically, in each iteration, we need to solve the following subproblems:
\begin{enumerate}
	\item Update all the auxiliary variables by closed-form solutions. 
	The complexity is given by $\mathcal{O} \left( \left( K+R+1 \right) N \right)$.
	
	\item Update the dual variable of dimension $K$ (with $K \ll N$) by solving the LP \eqref{60}, which can be easily solved using standard convex optimization algorithms. 
	The complexity of solving the LP \eqref{60} is approximately $\mathcal{O} \left( K^{3.5} N \right) $.
	
	\item Update the RIS reflecting coefficients by the closed-form solution \eqref{opt_x}. 
	The complexity is $\mathcal{O} \left( N \right)$.
\end{enumerate}
In summary, the overall computational complexity of the linear transform based algorithm is dominated by
\begin{align}
	\mathcal{O} \left( \mathcal{N} \left( K^2 + K+R+2 \right) N \right) ,
\end{align}
where $\mathcal{N}$ denotes the number of iterations.

The main advantage of the proposed linear transform based algorithm is its low computational complexity scaling with the number of RIS reflecting elements. 
Another notable advantage is that Algorithm~\ref{algorithm_2} is parameter-free, unlike the penalty-based algorithm, whose performance depends on the selection of the penalty coefficient.

\section{Extension to The Scenario Involving Unknown Fading Coefficient}
\label{Section_06}

In this section, we consider a more practical scenario where the fading coefficient $\alpha$ is unknown.
To tackle the problem in this scenario, we adopt the modified BCRLB proposed in \cite{1055879} as a performance metric for estimating $\eta$, which is given by
\begin{align} \label{69}
	\mathcal{B}(\mathbf{x})
	\triangleq
	\mathop{\mathbb{E}_{q \left( \alpha \right)}} \!
	\left[ 
		\frac{1}{\mathop{\mathbb{E}_{q \left( \eta \hspace{0.5pt}| \alpha \right)}} \left[ \mathsf{FI} \left( \eta \hspace{0.5pt}| \alpha \right) \right] + \mathsf{FIP} \left( \eta \hspace{0.5pt}| \alpha \right)}
	\right] ,
\end{align}
where $\mathsf{FI} \left( \eta \hspace{0.5pt}| \alpha \right)$ is the Fisher information of $\eta$ given a fixed $\alpha$ as previously defined in \eqref{FI} with a slight abuse of notation, the optimization variable $\mathbf{x}$ is embedded in $\mathsf{FI} \left( \eta \hspace{0.5pt}| \alpha \right)$, and the conditional prior Fisher information $\mathsf{FIP} \left( \eta \hspace{0.5pt}| \alpha \right)$ is given by
\begin{align}
	\mathsf{FIP} \left( \eta \hspace{0.5pt}| \alpha \right)
	=
	\mathop{\mathbb{E}_{q \left( \eta \hspace{0.5pt}| \alpha \right)}}\!
	\left[ 
	\left( \frac{\partial \ln q \left( \eta \hspace{0.5pt}| \alpha \right)}{\mathop{\partial} \eta} \right)^2
	\right] .
\end{align}
Then, the problem of optimally designing the RIS beamformer for joint sensing and communications can be formulated as
\begin{subequations} \label{76}
	\begin{align}
		\textbf{(P5):} \;\;\;
		\mathop{\mathrm{maximize}}_{\mathbf{x}} \;\;
		&
		- \mathcal{B}(\mathbf{x})  \\
		\mathop{\mathrm{subject\;to}} \;\;
		&
		\frac{\gamma_k(\mathbf{x})}{p_k} \geq \Gamma_k, \ \forall k, \\
		&
		\left| x_n \right| = 1, \ \forall n.
	\end{align}
\end{subequations}
If the distributions of $\alpha$ and $\eta$ are independent, i.e., $q \left( \eta \hspace{0.5pt}| \alpha \right) = q \left( \eta \right)$, which holds true at the initial sensing stage, the BCRLB can be rewritten as
\begin{align} \label{B}
	\mathcal{B}(\mathbf{x})
	=
	\mathop{\mathbb{E}_{q \left( \alpha \right)}} \!
	\left[ 
		\frac{1}{2 \hspace{1pt} p \hspace{-1pt} \left| \alpha \right|^2 \mathcal{A}(\mathbf{x}) + \mathsf{FIP} \left( \eta \right)}
	\right],
\end{align}
where $\mathcal{A}$ is defined in \eqref{11}.
It can be observed from the above that minimizing $\mathcal{B}$ is equivalent to maximizing $\mathcal{A}$ since $\mathcal{B}$ is a monotonically decreasing function of $\mathcal{A}$.
Therefore, solving problem (P5) is similar to solving problem (P1).

A more challenging case arises when the distributions of $\alpha$ and $\eta$ are dependent.
Different from \eqref{11}, it is quite challenging to extract the optimization variable $\mathbf{x}$ from the expectations in \eqref{69}, and Theorem~\ref{thm:second_moment} is no longer applicable.
To tackle this challenge, we employ uniform sampling for~approximating the expectations
and rewrite $\mathcal{B}$ as follows:
\begin{align}
	\label{eq:BCRB_average}
	\mathcal{B}(\mathbf{x})
	=
	\sum_{i=1}^{S}
	\frac{w_i}{
		\mathop{\displaystyle \sum_{j=1}^{L}} w_{i,j}^{\mathsf{o}} 
		\mathsf{FI} \left( \eta_j | \alpha_i \right)
		+ 
		\mathsf{FIP} \left( \eta \hspace{0.5pt}| \alpha_i \right)
	} ,
\end{align}
where $S$ and $L$ are the numbers of sampling points for $\alpha$ and $\eta$, respectively, the weights are $w_i = {q \left( \alpha_i \right)}/{\sum_{t=1}^{S} q \left( \alpha_t \right)} \geq 0$, $w_{i,j}^{\mathsf{o}} = {q \left( \eta_j | \alpha_i \right)}/{\sum_{t=1}^{L} q \left( \eta_t | \alpha_i \right)} \geq 0$, and $i$, $j$, and $t$ represent the indices of the samples.

The outer function in $-\mathcal{B}$ with respect to each $\mathsf{FI} \left( \eta_j | \alpha_i \right)$, which is a standard multi-dimensional ratio, is expressed as
\begin{align}
	\mathsf{h}^{\mathsf{o}}(\mathbf{\Psi}) 
	=
	- \sum_{i=1}^{S}
	\frac{w_i}{\mathop{\displaystyle \sum_{j=1}^{L}} w_{i,j}^{\mathsf{o}} \psi_{i,j} + \mathsf{FIP} \left( \eta \hspace{0.5pt}| \alpha_i \right)},
\end{align}
where $\psi_{i,j}$ denotes the $(i,j)$-th entry of the matrix $\mathbf{\Psi}$.
Unlike the previous scenario where $\alpha$ is fixed, this outer function is not nondecreasing over the entire $\mathbb{R}^{S \times L}$.
It can be verified that this function is nondecreasing in each $\psi_{i,j}$ over the following domain set:
\begin{align}
	\mathcal{H}^{\mathsf{o}}
	=
	\left\lbrace 
	\mathbf{\Psi} 
	\; \left| \
	\sum_{j=1}^{L} w_{i,j}^{\mathsf{o}} \psi_{i,j} + \mathsf{FIP} \left( \eta \hspace{0.5pt}| \alpha_i \right) > 0, \ \forall i 
	\right.
	\right\rbrace.
\end{align}
The above domain corresponds to having strictly positive denominator terms.
It can be seen that the image of $\mathcal{X} = \left\lbrace \mathbf{x} \mathop{|} \left| x_n \right| = 1, \forall n \right\rbrace$ under $\mathsf{FI} \left( \eta_j | \alpha_i \right)$ is a subset of $\mathcal{H}^{\mathsf{o}}$, i.e., condition \eqref{condition_01}, is satisfied.
This is because all the Fisher information terms $\mathsf{FI} \left( \eta_j | \alpha_i \right)$ and $\mathsf{FIP} \left( \eta \hspace{0.5pt}| \alpha_i \right)$ are nonnegative for all $\mathbf{x}$.

We now define a matrix $\tilde{\mathbf{F}}$ with the $(i,j)$-th entry as $\tilde{\mathsf{FI}} \left( \eta_j | \alpha_i \right)$.
Each of the entries is a quadratic transformed function of $\mathsf{FI} \left( \eta_j | \alpha_i \right)$ and is expressed as
\begin{align}
	& 
	\tilde{\mathsf{FI}} \left( \eta_j | \alpha_i \right)
	=  \notag \\
	&
	2 \mathop{\mathfrak{Re}} \! 
	\left\lbrace 
	(\boldsymbol{\lambda}_{i,j}^{\mathsf{o}})^{\dagger} \left( \sqrt{2p} \hspace{1pt} \alpha_i \dot{\mathbf{U}}(\eta_j) \mathbf{x} \right) 
	\right\rbrace
	-
	(\boldsymbol{\lambda}_{i,j}^{\mathsf{o}})^{\dagger} \mathbf{\Sigma}^{\mathsf{o}}(\mathbf{x}) \boldsymbol{\lambda}_{i,j}^{\mathsf{o}}.
\end{align}
Then, Theorem~\ref{thm:FCFP} can be applied to problem (P5), yielding 
\begin{subequations} \label{77}
	\begin{align}
		\mathop{\mathrm{maximize}}_{\mathbf{x}, \boldsymbol{\lambda}^{\mathsf{o}}_{i,j}, \boldsymbol{\lambda}^{\mathsf{c}}_k} \;\;
		&
		\mathsf{h}^{\mathsf{o}} \! \left( \tilde{\mathbf{F}} \right)   \label{69a} \\
		\mathop{\mathrm{subject\;to}} \;\;
		& \;
		\tilde{\mathsf{f}}^{\mathsf{c}}_k ( \mathbf{x}, \boldsymbol{\lambda}^{\mathsf{c}}_{k} ) \geq \Gamma_k,  \ \forall k,  \\
		& 
		\left| x_n \right| = 1 , \ \forall n ,  \\
		&\;
		\tilde{\mathbf{F}} \in \mathcal{H}^{\mathsf{o}},  \label{66b} 
	\end{align}
\end{subequations}
where the transformed constraint function $\tilde{\mathsf{f}}^{\mathsf{c}}_{k} (\mathbf{x}, \boldsymbol{\lambda}^{\mathsf{c}}_{k})$ is given in \eqref{sinrFP}.
The objective function \eqref{69a} is \emph{concave} over $\mathbf{x}$.

Then, utilizing the penalty-based approach in Section \ref{Section IV-B}, the problem \eqref{77} can be further transformed into
\begin{subequations}
	\begin{align}
		\mathop{\mathrm{maximize}}_{\mathbf{x}, \mathbf{z}, \boldsymbol{\lambda}^{\mathsf{o}}_{i,j}, \boldsymbol{\lambda}^{\mathsf{c}}_k} \;\;
		&\;
		\mathsf{h}^{\mathsf{o}} \! \left( \tilde{\mathbf{F}} \right) 
		- \mu \left\| \mathbf{x} - \mathbf{z} \right\|_2^2   \\
		\mathop{\mathrm{subject\;to}} \;\;
		& \;
		\tilde{\mathsf{f}}^{\mathsf{c}}_k ( \mathbf{x}, \boldsymbol{\lambda}^{\mathsf{c}}_{k} ) \geq \Gamma_k, \ \forall k,  \\
		& 
		\left| x_n \right| \leq 1 , \ \left| z_n \right| = 1, \ \forall n,  \\
		&
		\sum\limits_{j=1}^{L} 
		w_{i,j}^{\mathsf{o}}
		\tilde{\mathsf{FI}} \left( \eta_j | \alpha_i \right) + \mathsf{FIP} \left( \eta \hspace{0.5pt}| \alpha_i \right) > 0, \ \forall i.  \label{set_cons}
	\end{align}
\end{subequations}
Note that \eqref{set_cons} is equivalent to \eqref{66b}.
Similar to problem (\ref{49}), the above problem can be solved by iteratively optimizing the RIS beamforming vector and all the auxiliary variables.
The details are omitted here.
Note that the linear transform based algorithm is not applicable here due to the nested fractional structure of the objective function.

\section{Numerical Results}
\label{Section_07}

In this section, we provide simulation results to show the effectiveness of the proposed algorithms for RIS-assisted joint sensing and communication.
The simulation environment is set as follows.

\begin{itemize}
	\item
	The distance between the RIS and the BS is $100$ meters.
	The distance between the users and the RIS is $20$ meters.
	The pathloss exponent is $2$.
	The pathloss of the user-RIS link and the RIS-BS link are respectively given by
	\begin{align}
		& - 30 - 20 \log \left( 20 \right) \hspace{5pt} = - 56 \text{ dB},  \\
		& - 30 - 20 \log \left( 100 \right) = - 70 \text{ dB}.
	\end{align}

	\item
	The BS is equipped with $8$ antennas, which are arranged in a uniform linear array.
	The RIS is a planar with~$10 \times 10$ reflecting elements.
	Both the spacing between the antennas and the reflecting elements are half wavelength.

	\item
	The transmit power of each user is set to be $15$~dBm, the noise power at the BS is set as $-90$~dBm, and the SINR threshold is set to be $10$~dB.

	\item 
	We use the CVX with MOSEK as the optimization solver to handle the convex optimization problem involved in the optimization process.
\end{itemize}

\subsection{RIS Reflecting Beampatterns}

\begin{figure}[!t]
	\centering
	\vspace{-11pt}
	\includegraphics[width = 1 \linewidth]{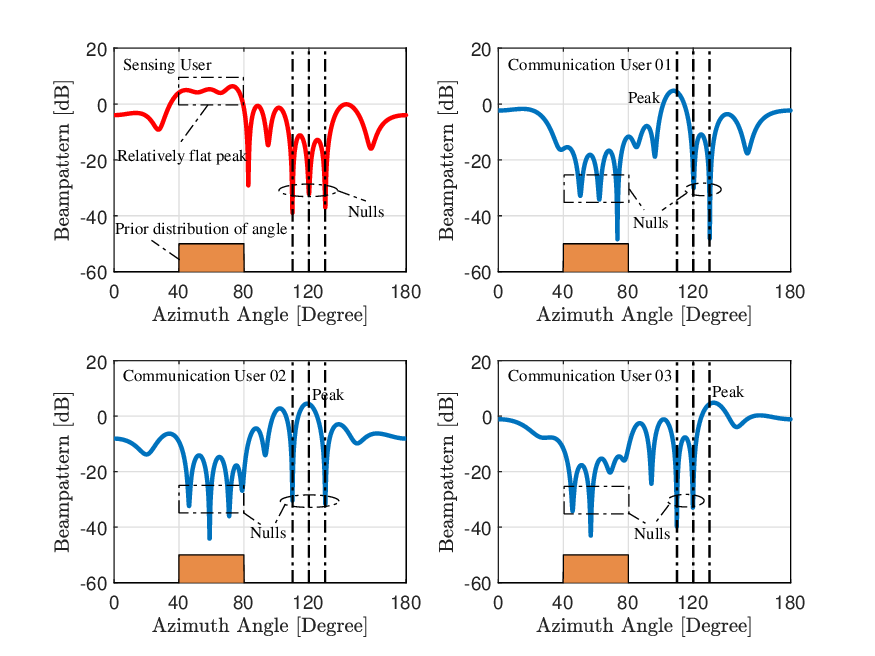}
	\caption{RIS reflecting beampatterns for the sensing and communication users with a uniform prior distribution for $\eta$ in the range $\left[ 40^{\circ}, 80^{\circ} \right]$.}
	\label{Fig_Beam}
\end{figure}

\begin{figure}[!t]
	\centering
	\includegraphics[width = 1 \linewidth]{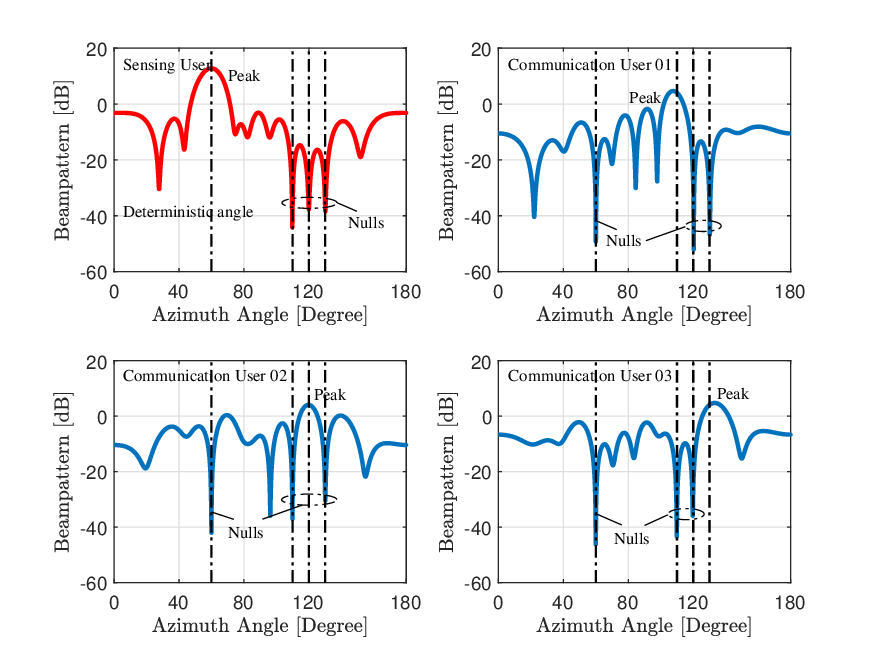}
	\caption{RIS reflecting beampatterns for the sensing and communication users with a deterministic $\eta$.}
	\label{Fig_Beam2}
\end{figure}

We use the beampatterns to illustrate the effectiveness of the proposed algorithms. 
This subsection deals with the case where $\alpha$ is known; the beampatterns for the case where $\alpha$ is unknown are similar and thus are omitted here.

The RIS reflecting beampattern for the $k$-th communication user is defined as follows:
\begin{align}
	\mathcal{Q}_k (\phi)
	=
	\left| 
		\mathbf{w}_k^{\dagger} 
		\mathbf{G} 
		\mathop{\mathrm{diag}} \{ \mathbf{x} \} 
		\mathbf{v}(\phi)
	\right|^2,
\end{align}
which is the received signal power after linear minimum mean square error (LMMSE) combining at the BS from a unit-power transmitter at an azimuth angle $\phi$ with respect to the RIS.

Although the receive combiner is generally not used for sensing, for comparison purposes, we also plot the received signal power for the sensing user using a linear combiner $\mathbf{w}$ similar for the communication users as given below:
\begin{align}
	\mathbf{w}
	= \boldsymbol{\delta}_{\max} \! \left( 
		\left( \mathbf{\Sigma}^{\mathsf{o}}(\mathbf{x}) \right)^{-1} 
		\mathop{\mathbb{E}_{q \left( \eta \right)}} \! \left[ 
			(\dot{\mathbf{U}} (\eta) \mathbf{x}) 
			(\dot{\mathbf{U}} (\eta) \mathbf{x})^{\dagger} 
		\right] 
	\right), 
\end{align}
where $\boldsymbol{\delta}_{\max} \left(\mathbf{A}\right)$ denotes the eigenvector corresponding to the largest eigenvalue of the matrix $\mathbf{A}$.
Specifically, we investigate the following two scenarios: 
\begin{enumerate}
	\item 
	The azimuth angle $\eta$ of the sensing user has a uniform prior distribution in the range $\left[40^{\circ}, 80^{\circ} \right]$.
	
	\item 
	The azimuth angle $\eta$ of the sensing user is deterministic at $60^{\circ}$.
\end{enumerate}
In both scenarios, there are three communication users located at $110^{\circ}$, $120^{\circ}$, and $130^{\circ}$, respectively.
Note that the proposed method is applicable to arbitrary prior distributions of the azimuth angle; the above two scenarios are chosen for illustrative purposes.

The designed RIS reflecting beampatterns have intuitive interpretations.
From Figs.~\ref{Fig_Beam} and \ref{Fig_Beam2}, one can observe that the solution of the beamforming optimization problem can produce beams that are aligned with the directions of communication users, while also having a peak matching the prior distribution of the sensing angle.
Moreover, the nulls of the beampatterns for the communication users are aligned with both the directions of interference and the directions corresponding to the prior distribution of the sensing azimuth angle.
These aligned peaks and nulls help focus power in the desired directions while reducing interference, thereby improving the performance of joint sensing and communications.

\subsection{Posterior Distributions of Estimated Parameters}

\begin{figure}[!t]
	\centering
	\vspace{-6pt}
	\includegraphics[width = 1 \linewidth]{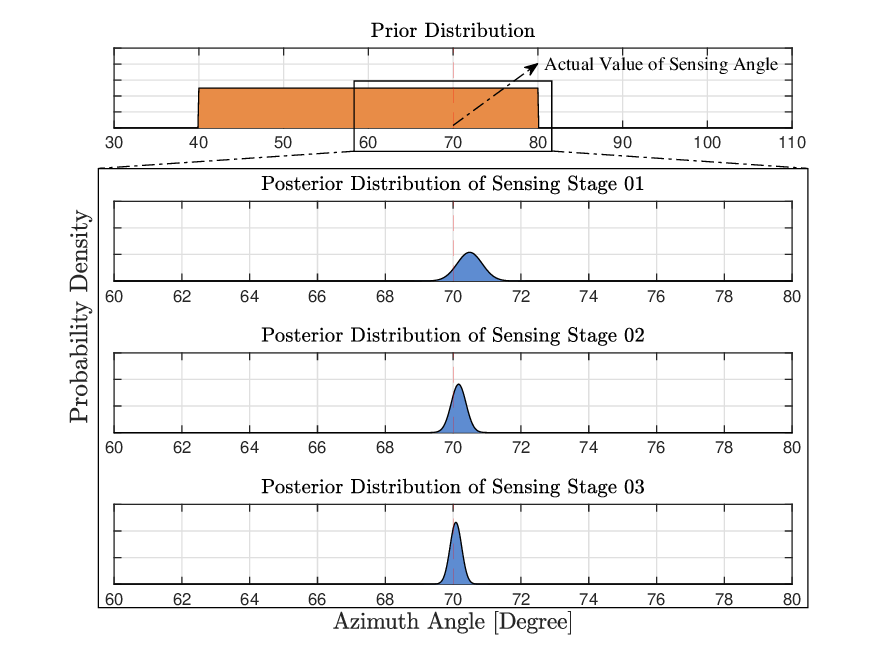}
	\caption{Posterior distributions of the angle $\eta$ over three iterations with adaptive RIS sensing (fading coefficient $\alpha$ is known).}
	\label{Fig_04}
\end{figure}

\begin{figure}[!t]
	\centering
	\includegraphics[width = 1 \linewidth]{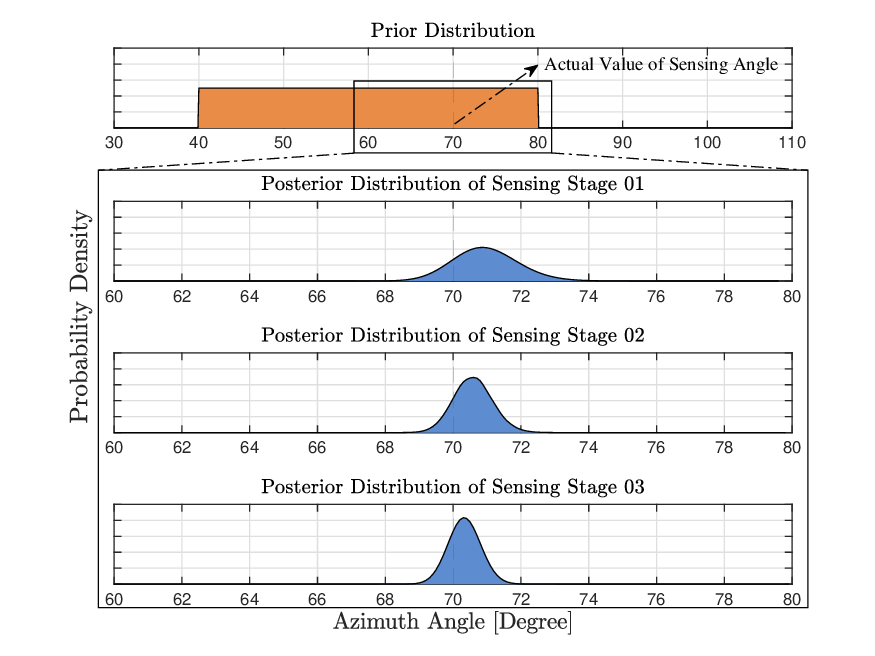}
	\caption{Posterior distributions of the angle $\eta$ over three iterations with adaptive RIS sensing (fading coefficient $\alpha$ is unknown).}
	\label{Fig_05}
\end{figure}

\begin{figure}[!t]
	\centering
	\vspace{-10pt}
	\includegraphics[width = 1 \linewidth]{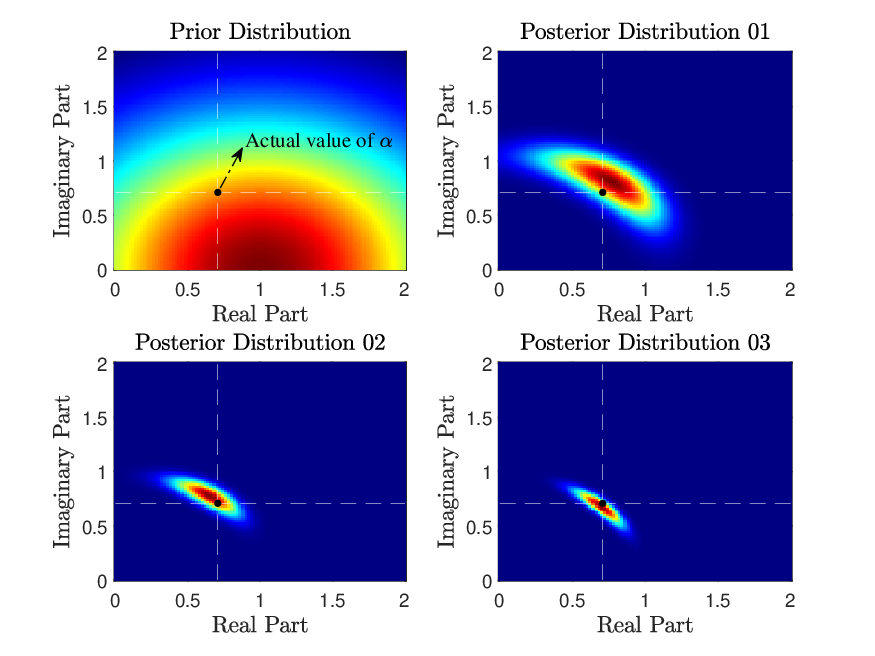}
	\caption{Posterior distributions of the fading coefficient $\alpha$ over three iterations with adaptive RIS sensing.}
	\label{Fig_06}
\end{figure}

To further illustrate the sensing performance of the proposed algorithms, we show the evolution of the posterior distributions after several iterations of sensing stages in Figs.~\ref{Fig_04} -- \ref{Fig_06}.
First, we consider the case where $\alpha$ is known.
The posterior probability function of the $(t+1)$-th sensing stage is computed as follows.
Note that the communication signals are regarded as noise for sensing. 
In this case, we have
\begin{align} \label{posteriorD}
	& 
	\mathsf{Pr} \left( \eta \mathop{|} \mathbf{y} [t+1] \right) 
	\propto
	\mathsf{Pr} \left( \eta \mathop{|} \mathbf{y} [t] \right) 
	\mathop{\boldsymbol{\cdot}}
	\mathcal{L} \left( \mathbf{y} [t+1]; \eta \right)  \notag  \\
	& 
	= \mathsf{Pr} \left( \eta \mathop{|} \mathbf{y} [t] \right) \mathop{\boldsymbol{\cdot}}  \notag  \\
	& \hspace{13.5pt}
	\mathcal{CN} \Big(\mathsf{mean} \left( \mathbf{y} [t+1] \mathop{|} \eta \right), \mathsf{Cov} \left( \mathbf{y} [t+1] \mathop{|} \eta \right) \Big) ,
\end{align}
where $\mathcal{L} \left( \mathbf{y} [t+1]; \eta \right)$ is the Gaussian likelihood function with the mean and covariance matrix given in (\ref{mean}) and (\ref{cov}), respectively, and 
$\mathsf{Pr} \left( \eta \mathop{|} \mathbf{y} [t] \right)$ denotes the posterior distribution from the previous iteration, which is adopted as the prior distribution $q \left( \eta \right)$ for the current iteration.
Then, the above process can be repeated until the accuracy satisfies the requirement.

We show the results corresponding to the scenario of Fig.~\ref{Fig_Beam}, where the fading coefficient $\alpha$ is known, and the azimuth angle $\eta$ has a uniform distribution in the range $\left[40^{\circ}, 80^{\circ} \right]$.
The actual angle of the sensing user is set to be $70^{\circ}$.
The communication signals are randomly generated from a Gaussian distribution.
The pilot signal of the sensing user is set as a known sequence with transmit power $p = 15$~dBm. 
In Fig.~\ref{Fig_04}, we illustrate the posterior distributions of $\eta$ after three iterations using the RIS reflection coefficients optimized by the proposed quadratic transform based algorithm.
From Fig.~\ref{Fig_04}, one can observe that as the number of iterations increases, the posterior distribution of $\eta$ rapidly converges to a highly concentrated distribution with a peak at the true sensing angle.
This validates that the proposed beamforming design is highly effective for sensing.

In Figs.~\ref{Fig_05} and \ref{Fig_06}, we plot the results for the scenario where the fading coefficient $\alpha$ is unknown.
The prior distribution of $\alpha$~is set as $\left(-56 \text{ dB}\right) \times \mathcal{CN} \left( 1, 1 \right)$, and the actual value of $\alpha$ is set as $\left(-56 \text{ dB}\right) \times \left( 0.7 + 0.7 j \right)$.
In this case, the posterior distribution is modified as follows:
\begin{align} \label{posterior}
	& 
	\mathsf{Pr} \left( \eta, \alpha \mathop{|} \mathbf{y} [t+1] \right) 
	\propto
	\mathsf{Pr} \left( \eta, \alpha \mathop{|} \mathbf{y} [t] \right) 
	\mathop{\boldsymbol{\cdot}}
	\mathcal{L} \left( \mathbf{y} [t+1]; \eta, \alpha \right) \notag  \\
	&
	= \mathsf{Pr} \left( \eta, \alpha \mathop{|} \mathbf{y} [t] \right) 
	\mathop{\boldsymbol{\cdot}}  \notag  \\
	& \hspace{13.5pt}
	\mathcal{CN} \Big(\mathsf{mean} \left( \mathbf{y} [t+1] \mathop{|} \eta, \alpha \right), \mathsf{Cov} \left( \mathbf{y} [t+1] \mathop{|} \eta, \alpha \right) \Big) ,
\end{align}
and the posterior distributions of $\alpha$ and $\eta$ can be computed by $\int_{\eta} \mathsf{Pr} \left( \eta, \alpha \mathop{|} \mathbf{y}[t+1] \right) d \hspace{1pt} \eta$ and $\int_{\alpha} \mathsf{Pr} \left( \eta, \alpha \mathop{|} \mathbf{y}[t+1] \right) d \hspace{1pt} \alpha$, respectively.
It can be observed from Figs.~\ref{Fig_05} and \ref{Fig_06} that with increasing number of iterations, both the posterior distributions of $\alpha$ and $\eta$ rapidly converge to a concentrated distribution.
This shows that the proposed beamforming design is highly effective for sensing.
However, compared to Fig.~\ref{Fig_04}, i.e., the scenario where $\alpha$ is already known, the convergence rate in Fig.~\ref{Fig_05} is slower, reflecting the impact of uncertainty in the fading coefficient~on the azimuth angle estimation.

\subsection{Estimation and Optimization Performance}

\begin{figure}[!t]
	\centering
	\vspace{-10pt}
	\includegraphics[width = 1 \linewidth]{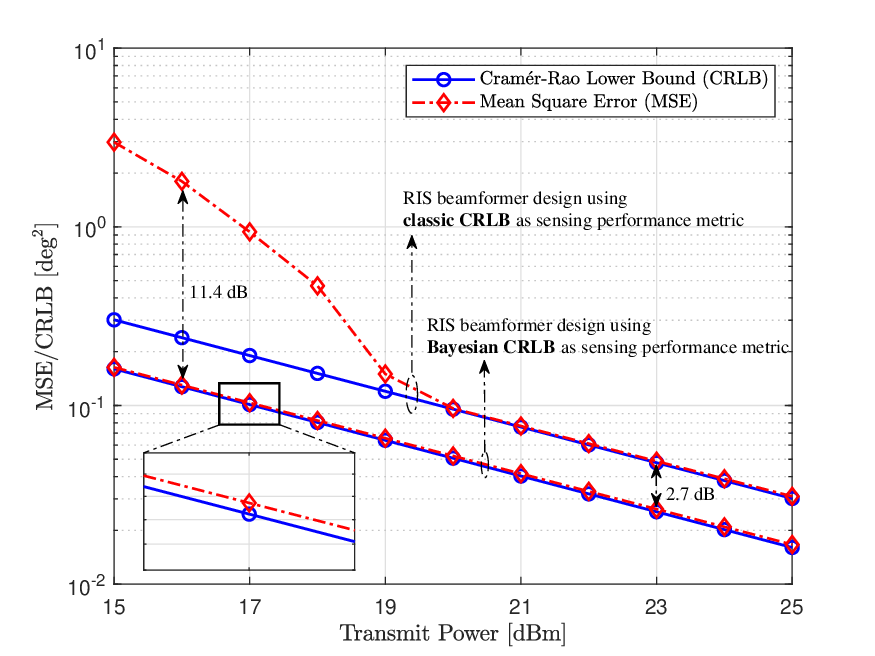}
	\caption{MSE and CRLB of estimating the azimuth angle of the sensing user ($\eta = 70^{\circ}$) with different RIS beamformers (fading coefficient $\alpha$ is known).}
	\label{Fig_07}
\end{figure}

We now show that better performance can be achieved using the BCRLB as the sensing performance metric than using the classic CRLB.
Consider the scenario of Fig.~\ref{Fig_Beam}, i.e., the azimuth angle of sensing user is uniformly distributed in the range of $\left[40^{\circ}, 80^{\circ} \right]$, and the actual azimuth angle of the sensing user is set at $70^{\circ}$.
The classic CRLB computation relies on the actual value of $\eta$, which is unknown and needs to be estimated.
One possible approach is to utilize the mean of $\eta$ to compute the CRLB.
In contrast, the BCRLB leverages the prior distribution of $\eta$. 
For both cases, maximum likelihood estimation (MLE)~is used to estimate the azimuth angle of the sensing user, i.e.,
\begin{align}
	\hat{\eta}
	& =
	\mathop{\arg\max}_{\eta} \;
	\mathcal{L} \left( \mathbf{y}; \eta \right), \\
	\left( \hat{\eta}, \hat{\alpha} \right) 
	& =
	\mathop{\arg\max}_{\eta, \alpha} \;
	\mathcal{L} \left( \mathbf{y}; \eta, \alpha \right) .
\end{align}
Note that although the proposed active sensing scheme allows for the incorporation of prior knowledge through maximum a posteriori (MAP) estimation, we opt for MLE here to facilitate a theoretically consistent comparison with the CRLB, which assumes unbiased estimators.
The MSE of the estimation of $\eta$ is obtained through $10,000$ Monte-Carlo trials with randomly generated noise and communication data symbols.

\begin{figure}[!t]
	\centering
	\includegraphics[width = 1 \linewidth]{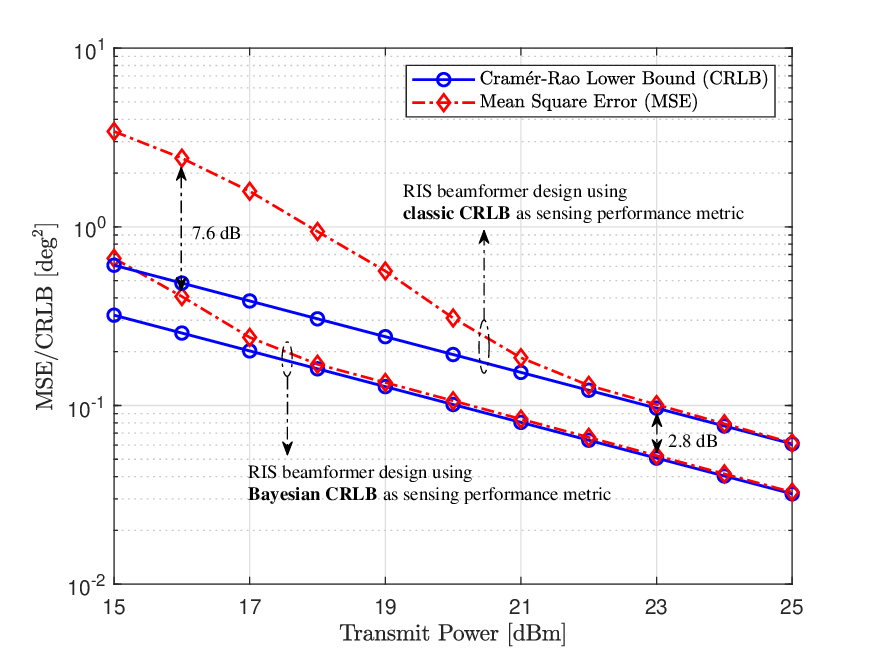}
	\caption{MSE and CRLB of estimating the azimuth angle of the sensing user ($\eta = 70^{\circ}$) with different RIS beamformers (fading coefficient $\alpha$ is unknown).}
	\label{Fig_08}
\end{figure}

\begin{table}[!t]
	\centering
	\caption{Runtime for FCFP with different variable dimensions.}
	\renewcommand{\arraystretch}{1.5}
	\label{TableI}
	\begin{tabular}{c c c c}
		\toprule
		& 
		\multicolumn{1}{p{2cm}}{\vspace{-12pt}\textbf{Runtime \;\;\;\;\;\;\; per Iteration [s]}} 
		& 
		\multicolumn{1}{p{2cm}}{\vspace{-12pt}\textbf{\textbf{No. of Iterations to Convergence}}} 
		& 
		\multicolumn{1}{p{2cm}}{\vspace{-12pt}\textbf{Runtime to \;\;\;\;\;\;\;\; Convergence [s]}}  \\ 
		
		\hline\hline
		\multicolumn{1}{c}{} 
		&
		\multicolumn{3}{l}{Number of reflecting elements: $N = 10 \times 10 = 100$}  \\ 
		\hline
		\multicolumn{1}{c|}{\textbf{PN-QT}}    
		&0.2746  &146   &40.09  \\
		\multicolumn{1}{c|}{\textbf{CM-LT}}  
		&0.0042  &2658  &11.14  \\  
		
		\hline\hline
		\multicolumn{1}{c}{} 
		&
		\multicolumn{3}{l}{Number of reflecting elements: $N = 15 \times 15 = 225$}  \\ 
		\hline
		\multicolumn{1}{c|}{\textbf{PN-QT}}    
		&0.6174  &229   &141.38  \\
		\multicolumn{1}{c|}{\textbf{CM-LT}}  
		&0.0241  &2163  &52.08   \\  
		
		\hline\hline
		\multicolumn{1}{c}{} 
		& 
		\multicolumn{3}{l}{Number of reflecting elements: $N = 20 \times 20 = 400$}  \\ 
		\hline
		\multicolumn{1}{c|}{\textbf{PN-QT}}    
		&2.4513  &394   &965.81  \\
		\multicolumn{1}{c|}{\textbf{CM-LT}}  
		&0.1259  &3093  &389.68  \\ 
		
		\hline\hline
	\end{tabular}
\end{table}

In Fig.~\ref{Fig_07}, we present the MSE and the CRLB of estimating the actual azimuth angle $\eta = 70^{\circ}$ with different RIS reflection coefficients designed by optimizing different sensing metrics.
One can observe from Fig.~\ref{Fig_07} that adopting the BCRLB as the sensing performance metric can provide much better sensing performance than that of using the classic CRLB, particularly in the low transmit power regime.
The estimation performance of adopting the classic CRLB as the metric is more sensitive to the power of received signals than using the BCRLB.
Thus, the CRLB is not achievable using the MLE in the low transmit power regime for the case of using the classic CRLB as metric, making the performance gap relative to the case of using the BCRLB particularly large in the low transmit power regime.

In Fig.~\ref{Fig_08}, we consider the same setup with Fig.~\ref{Fig_07}, but with an unknown fading coefficient $\alpha$.
The prior distribution of $\alpha$ is set as $\left(-56 \text{ dB}\right) \times \mathcal{CN} \left( 1, 1 \right)$, and the actual value of $\alpha$ is set as $\left(-56 \text{ dB}\right) \times \left( 0.7 + 0.7 j \right)$.
From Fig.~\ref{Fig_08}, in addition to observing results similar to those in Fig.~\ref{Fig_07}, it can be observed that both the MSE and the CRLB are higher than those in Fig.~\ref{Fig_07}, reflecting the impact of uncertainty in $\alpha$ on the angle estimation.
It can also be observed that the uncertainty in $\alpha$ makes the estimation performance of both cases of using the classic CRLB and the BCRLB more sensitive to the received signal power.

\begin{figure}[!t]
	\centering
	\includegraphics[width = 1 \linewidth]{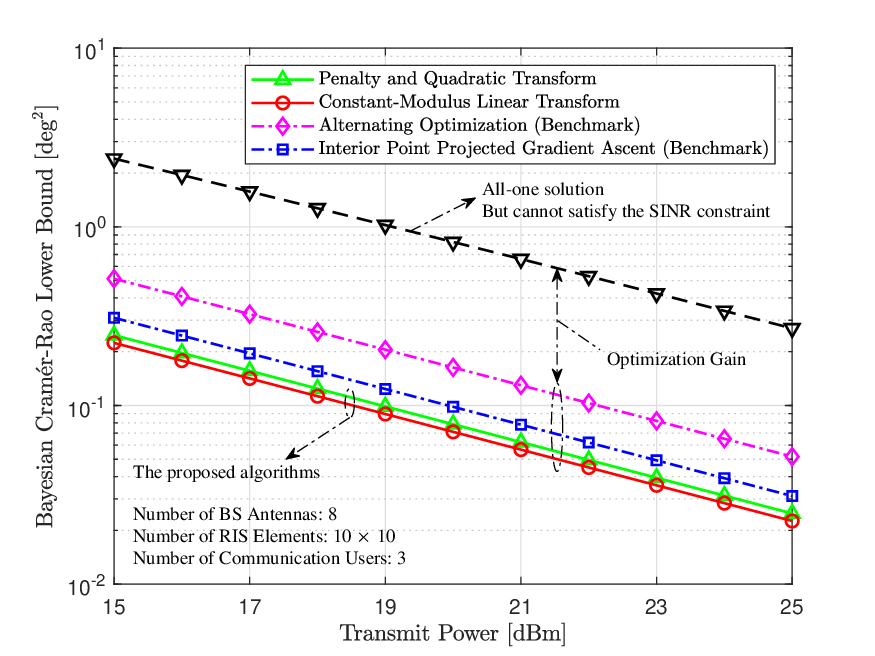}
	\caption{Bayesian Cram\'{e}r-Rao lower bound versus the transmit power using different optimization algorithms.}
	\label{Fig_09}
\end{figure}

\begin{table}[!t]
	\centering
	\caption{Optimized Bayesian Cram\'{e}r-Rao Lower Bound [$\mathrm{deg}^2$] for Different Algorithms in Different Scenarios.}
	\renewcommand{\arraystretch}{1.5}
	\label{TableII}
	\begin{tabular}{p{1.75cm} p{1.75cm} p{1.75cm} p{1.75cm}}
		\toprule
		\multicolumn{1}{p{1.75cm}}{\textbf{CM-LT}} 
		& \multicolumn{1}{p{1.75cm}}{\textbf{PN-QT}} 
		& \multicolumn{1}{p{1.75cm}}{\textbf{AO}} 
		& \multicolumn{1}{p{1.75cm}}{\textbf{IPGA}}  \\ 
		
		\hline\hline
		\multicolumn{4}{l}{\textbf{Scenario 1}: Two communication users at $110^{\circ}$ and $130^{\circ}$.}  \\ 
		\hline
		0.205 &0.223  &0.384  &0.276  \\
		
		\hline\hline
		\multicolumn{4}{l}{\textbf{Scenario 2}: Three communication users at $110^{\circ}$, $120^{\circ}$, and $130^{\circ}$.}  \\ 
		\hline
		0.224  &0.247  &0.513  &0.310  \\
		
		\hline\hline
		\multicolumn{4}{l}{\textbf{Scenario 3}: Four communication users at $100^{\circ}$, $110^{\circ}$, $120^{\circ}$, and $130^{\circ}$.}  \\ 
		\hline
		0.230  &0.274  &0.819  &0.361  \\
		
		\hline\hline
	\end{tabular}
\end{table}

Next, we show the efficiency of the two proposed algorithms and validate the corresponding complexity analysis,~especially when the number of reflecting elements is large.
Table~\ref{TableI} shows the runtime per iteration, the number of iterations to converge, and the total convergence time of the two proposed algorithms, referred to as PN-QT for the penalty and quadratic transform based algorithm and CM-LT for the constant-modulus linear transform based algorithm, under the same setup. 
From the table, one can observe that the linear transform-based algorithm has a significantly shorter per-iteration runtime compared with the quadratic transform-based algorithm, which aligns with the prior complexity analysis.
Although the linear transform-based algorithm requires more iterations to converge, its overall run-time remains lower than that of the quadratic transform-based algorithm,~especially when $N$ is large.

We now present the optimization performance of the proposed algorithms.
To the best of our knowledge, no existing work has dealt with the BS and RIS beamforming problem considered in this paper.
Thus, as a baseline, we adopt an interior-point-like benchmark to handle the SINR constraints and then use the projected gradient ascent method to solve the following problem:
\begin{equation}
	\mathop{\mathrm{maximize}}_{\left| x_n \right| = 1}
	\
	\mathcal{A}(\mathbf{x})  
	+ 
	\sum_{k=1}^{K} \frac{1}{\mu} \log \left( \gamma_k (\mathbf{x}) - \Gamma \right).
\end{equation}
The above benchmark is referred to as IPGA.
Alternatively, it is possible to optimize the RIS beamformer by first quantizing the phase shifts into $8$-bit levels, which provides a sufficiently fine resolution without noticeable performance degradation, followed by using the alternating optimization (AO) algorithm.
Fig.~\ref{Fig_09} plots the optimized BCRLB versus the transmit power.
One can observe that the proposed algorithms perform better than the two benchmark algorithms, which demonstrates the priority of the proposed algorithms.

Next, we show results for three scenarios: 
1) two communication users at $110^{\circ}$ and $130^{\circ}$; 2) three communication users at $110^{\circ}$, $120^{\circ}$, and $130^{\circ}$; 3) four communication users at $100^{\circ}$, $110^{\circ}$, $120^{\circ}$, and $130^{\circ}$.
As shown in Table~\ref{TableII}, the proposed algorithms exhibit better performance and adaptability to varying numbers and spatial configurations of communication users.

\begin{figure}[!t]
	\centering
	\vspace{-10pt}
	\includegraphics[width = 1 \linewidth]{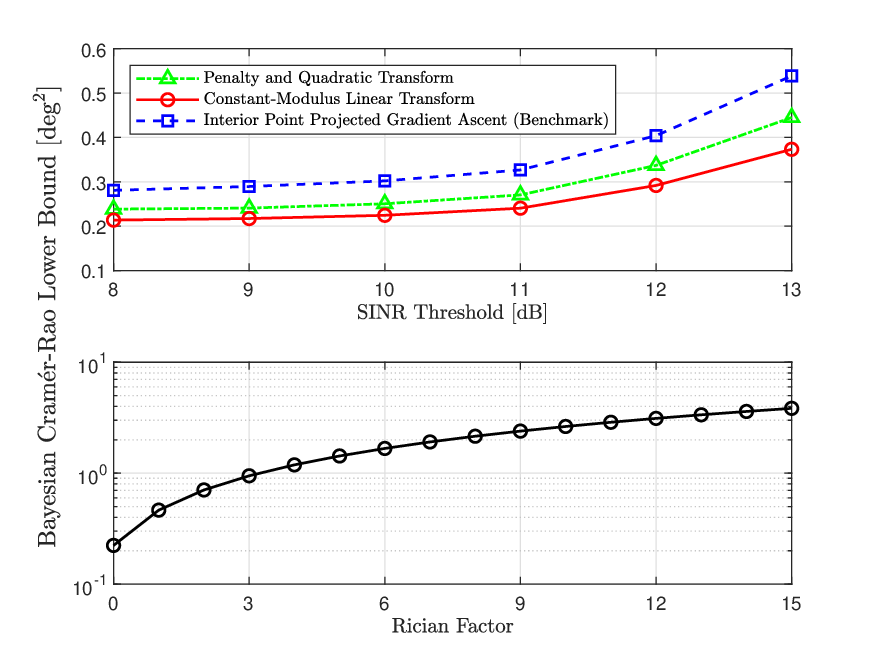}
	\caption{Bayesian Cram\'{e}r-Rao lower bound versus the SINR threshold and versus the Rician factor of the channel.}
	\label{Fig_10}
\end{figure}

Note that all the previous results use a fixed SINR threshold.
It is also important to show the impact of varying SINR targets on the sensing performance.
In Fig.~\ref{Fig_10}, we plot he BCRLB versus the SINR threshold.
As the SINR threshold increases, the BCRLB for sensing also increases, and the rate of increase becomes faster with higher SINR thresholds.
This is not only due to that the high SINR for communications can cause high interference for sensing signals, but also because the RIS needs to allocate more reflection resources for communication users to meet the higher SINR requirement.
Consequently, the sensing user is allocated fewer beamforming resources, resulting in worse estimation performance.

Finally we consider a more practical scenario where scattering exists in the wireless environment.
The corresponding channel model is refined to a Rician fading model as
\begin{align}
	\mathbf{u} 
	= \alpha \left( \sqrt{\frac{1}{1+\zeta}} \mathbf{U} ( \eta ) \mathbf{x} 
	+ \sqrt{\frac{\zeta}{1+\zeta}} \mathbf{G} \mathop{\mathrm{diag}} \{ \tilde{\mathbf{v}} \} \mathbf{x}  \right) ,
\end{align}
where $\zeta$ represents the Rician factor, and the elements of $\tilde{\mathbf{v}}$ are independently generated from $\mathcal{CN} \left( 0, 1 \right)$.
Since the NLoS part contains no angular information, it serves as noise to the angle estimation.
Here, $\zeta$ determines the scattering degree of the wireless propagation environment, and when $\zeta = 0$, the channel has only an LoS path as given in (\ref{los_channel}).
We plot the curves of the BCRLB versus the Rician factor in Fig.~\ref{Fig_10}.
One can observe that the BCRLB of estimation increases with the Rician factor, indicating that the sensing performance degrades as the richness of the scattering environment increases. 
This suggests that more complex environments require the use of additional resources, such as a larger number of antennas, more RIS reflecting elements, or higher transmit power, to maintain satisfactory performance.

\section{Conclusion}
\label{Section_08}

This paper proposes a methodology for RIS beamforming pattern design for
an uplink RIS-assisted integrated sensing and communications scenario. 
We formulate the optimization problem of minimizing the BCRLB of azimuth angle estimation for the sensing user, while imposing SINR constraints for multiple communication users.  
This problem is a non-convex fractional program with fractional constraints. 
We utilize two transform techniques to deal with the fractional structures in both the objective function and the constraints, which turn the problem into a sequence of convex subproblems. 
Simulation results demonstrate highly effective RIS beampattern design for both sensing and communications.

\appendices

\section{Proof of Theorem \ref{thm:second_moment}}
\label{app:theorem_1}

Since $\mathbf{\Sigma}^{\mathsf{o}}(\mathbf{x})$ is a positive definite matrix, we can decompose its inverse as
$(\mathbf{\Sigma}^{\mathsf{o}}(\mathbf{x}))^{-1} = \mathbf{\Upsilon}^{\dagger} \mathbf{\Upsilon}$.
According to the properties of the Kronecker product,
given matrices $\mathbf{A}$, $\mathbf{B}$, $\mathbf{C}$, and $\mathbf{D}$, we have
\begin{align}
	{\mathop{\mathrm{vec}}} \left( \mathbf{A} \mathbf{C} \mathbf{B} \right)
	& =
	\left( \mathbf{B}^{\mathsf{T}} \otimes \mathbf{A} \right) {\mathop{\mathrm{vec}}} \left( \mathbf{C} \right) ,  \\
	\left( \mathbf{A} \otimes \mathbf{B} \right) \left( \mathbf{C} \otimes \mathbf{D} \right)
	& =
	\left( \mathbf{A} \mathbf{C} \right) \otimes  \left( \mathbf{B} \otimes \mathbf{D} \right).
\end{align}
Then, the inner part of \eqref{11} can be rewritten as follows:
\begin{align} \label{Kronecker}
	( & \dot{\mathbf{U}} ( \eta ) \mathbf{x} )^{\dagger}
	(\mathbf{\Sigma}^{\mathsf{o}}(\mathbf{x}))^{-1}
	( \dot{\mathbf{U}} ( \eta ) \mathbf{x} )
	= 
	( \mathbf{\Upsilon} \dot{\mathbf{U}} ( \eta ) \mathbf{x} )^{\dagger}
	( \mathbf{\Upsilon} \dot{\mathbf{U}} ( \eta ) \mathbf{x} )  \notag  \\
	&
	=
	( ( \mathbf{x}^{\mathsf{T}} \otimes \mathbf{\Upsilon} ) \mathop{\mathrm{vec}} (  \dot{\mathbf{U}} ( \eta ) ) )^{\dagger}
	( ( \mathbf{x}^{\mathsf{T}} \otimes \mathbf{\Upsilon} ) \mathop{\mathrm{vec}} (  \dot{\mathbf{U}} ( \eta ) ) )  \notag  \\
	&
	=
	{\mathop{\mathrm{vec}}}^{\dagger} ( \dot{\mathbf{U}} ( \eta ) )
	( \mathbf{x}^{*} \otimes \mathbf{\Upsilon}^{\dagger} )
	( \mathbf{x}^{\mathsf{T}} \otimes \mathbf{\Upsilon} )
	{\mathop{\mathrm{vec}}} ( \dot{\mathbf{U}} ( \eta ) )  \notag  \\
	&
	=
	{\mathop{\mathrm{vec}}}^{\dagger} ( \dot{\mathbf{U}} ( \eta ) )
	( \mathbf{x}^{*} \mathbf{x}^{\mathsf{T}} \otimes \mathbf{\Upsilon}^{\dagger} \mathbf{\Upsilon} )
	{\mathop{\mathrm{vec}}} ( \dot{\mathbf{U}} ( \eta ) )  \notag  \\
	&
	=
	\mathrm{\mathop{Tr}} ( ( \mathbf{x}^{*} \mathbf{x}^{\mathsf{T}} \otimes (\mathbf{\Sigma}^{\mathsf{o}}(\mathbf{x}))^{-1} ) ( {\mathop{\mathrm{vec}}} ( \dot{\mathbf{U}} ( \eta ) ) {\mathop{\mathrm{vec}}}^{\dagger} ( \dot{\mathbf{U}} ( \eta ) ) ) )  .   
\end{align}
Thus, the expectation in \eqref{11} can be equivalently  rewritten as
\begin{align} \label{newE}
	\mathcal{A}(\mathbf{x})
	= \mathrm{\mathop{Tr}} \left( \left( \mathbf{x}^{*} \mathbf{x}^{\mathsf{T}} \otimes (\mathbf{\Sigma}^{\mathsf{o}}(\mathbf{x}))^{-1} \right) 
	\dot{\mathbf{R}} \right) ,
\end{align}
where $\dot{\mathbf{R}}$ is the expectation given in \eqref{U}.
Since $\dot{\mathbf{R}}$ is a positive semidefinite matrix, it has an eigenvalue decomposition as
\begin{equation} \label{eigendecomposition}
	\dot{\mathbf{R}}
	= \sum_{r=1}^{R} \kappa_r \mathbf{r}_{r} \mathbf{r}_{r}^{\dagger},
\end{equation}
where $\kappa_r$ and $\mathbf{r}_{r}$ are the $r$-th eigenvalue and the corresponding eigenvector of $\dot{\mathbf{R}}$, respectively.
Substituting \eqref{eigendecomposition} into \eqref{newE} and applying the derivation in \eqref{Kronecker} in reverse give us \eqref{obj_Approx}.

\section{Proof of Theorem \ref{thm:FCFP}}
\label{app:theorem_2}

Based on Lemma~\ref{lem:quadratic_transform}, we have the following identities:
\begin{align} 
	& 
	\max_{\boldsymbol{\lambda}^{\mathsf{o}}_i} \; \tilde{\mathsf{f}}^{\mathsf{o}}_i (\mathbf{x}, \boldsymbol{\lambda}^{\mathsf{o}}_i)
	=
	\mathsf{f}^{\mathsf{o}}_i (\mathbf{x}),  \label{equivalence_01}  \\
	& 
	\max_{\boldsymbol{\lambda}^{\mathsf{c}}_{k,j}} \; \tilde{\mathsf{f}}^{\mathsf{c}}_{k,j} (\mathbf{x}, \boldsymbol{\lambda}^{\mathsf{c}}_{k,j})
	=
	\mathsf{f}^{\mathsf{c}}_{k,j} (\mathbf{x}).  \label{equivalence_02}
\end{align}
Then, conditions \eqref{condition_01} and \eqref{condition_02} give the following relations:
\begin{align} \label{87}
	&
	\left\lbrace 
		\left. \left(  
			\max_{\boldsymbol{\lambda}^{\mathsf{o}}_1} \; \tilde{\mathsf{f}}^{\mathsf{o}}_1 (\mathbf{x}, \boldsymbol{\lambda}^{\mathsf{o}}_1), 
			\ldots,
			\max_{\boldsymbol{\lambda}^{\mathsf{o}}_I} \; \tilde{\mathsf{f}}^{\mathsf{o}}_I (\mathbf{x}, \boldsymbol{\lambda}^{\mathsf{o}}_I) 
		\right) \right|
		\mathbf{x} \in \mathcal{X}		
	\right\rbrace  \notag \hspace{30pt} \\
	& \quad =
	\left\lbrace 
		\left( 
			\mathsf{f}^{\mathsf{o}}_1(\mathbf{x}), 
			\ldots, 
			\mathsf{f}^{\mathsf{o}}_I(\mathbf{x}) 
		\right) \mathop{|}
		\mathbf{x} \in \mathcal{X}
	\right\rbrace
	\subseteq
	\mathcal{H}^{\mathsf{o}},
\end{align}
and
\begin{align} \label{88}
	&
	\left\lbrace 
	\left. \left( 
		\max_{\boldsymbol{\lambda}^{\mathsf{c}}_{k,1}} \; \tilde{\mathsf{f}}^{\mathsf{c}}_{k,1} ( \mathbf{x}, \boldsymbol{\lambda}^{\mathsf{c}}_{k,1}), 
		\ldots, 
		\max_{\boldsymbol{\lambda}^{\mathsf{c}}_{k,J}} \; \tilde{\mathsf{f}}^{\mathsf{c}}_{k,J} ( \mathbf{x}, \boldsymbol{\lambda}^{\mathsf{c}}_{k,J}) 
	\right) \right|
	\mathbf{x} \in \mathcal{X}		
	\right\rbrace  \notag  \\
	& \quad =
	\left\lbrace 
		\left( 
			\mathsf{f}^{\mathsf{c}}_{k,1}(\mathbf{x}), 
			\ldots, 
			\mathsf{f}^{\mathsf{c}}_{k,J}(\mathbf{x}) 
		\right): 
		\mathbf{x} \in \mathcal{X} 
	\right\rbrace
	\subseteq
	\mathcal{H}_k^{\mathsf{c}}.
\end{align}
Based on the above relations, we can now prove Theorem~\ref{thm:FCFP}.
To this end, we first show that problem \eqref{equivalent Ray} is equivalent to
\begin{subequations} \label{81}
	\begin{align}
		\mathop{\mathrm{maximize}}_{\mathbf{x} \in \mathcal{X}} \;
		&\;
		\mathsf{h}^{\mathsf{o}} \! \left(  
			\max_{\boldsymbol{\lambda}^{\mathsf{o}}_1} \; \tilde{\mathsf{f}}^{\mathsf{o}}_1 (\mathbf{x}, \boldsymbol{\lambda}^{\mathsf{o}}_1), 
			\ldots, 
			\max_{\boldsymbol{\lambda}^{\mathsf{o}}_I} \; \tilde{\mathsf{f}}^{\mathsf{o}}_I (\mathbf{x}, \boldsymbol{\lambda}^{\mathsf{o}}_I) 
		\right)   \\
		\mathop{\mathrm{subject\;to}} \;
		&\;
		\mathsf{h}^{\mathsf{c}}_k \! \left( 
			\max_{\boldsymbol{\lambda}^{\mathsf{c}}_{k,1}} \; \tilde{\mathsf{f}}^{\mathsf{c}}_{k,1} ( \mathbf{x}, \boldsymbol{\lambda}^{\mathsf{c}}_{k,1} ), 
			\ldots, 
			\max_{\boldsymbol{\lambda}^{\mathsf{c}}_{k,J}} \; \tilde{\mathsf{f}}^{\mathsf{c}}_{k,J} ( \mathbf{x}, \boldsymbol{\lambda}^{\mathsf{c}}_{k,J} ) 
		\right)  \notag  \\
		& 
		\geq c_k,  \  \forall k. \label{81b}
	\end{align}
\end{subequations}
To see why, first note that 
\begin{subequations} \label{90}
	\begin{align}
		&\hspace{2pt} \mathop{\mathrm{max}}_{\mathbf{x} \in \mathcal{X}, \boldsymbol{\lambda}^{\mathsf{o}}_i, \boldsymbol{\lambda}^{\mathsf{c}}_j} \;
		\mathsf{h}^{\mathsf{o}} \! \left( \tilde{\mathsf{f}}^{\mathsf{o}}_1 ( \mathbf{x}, \boldsymbol{\lambda}^{\mathsf{o}}_1 ), \ldots, \tilde{\mathsf{f}}^{\mathsf{o}}_I ( \mathbf{x}, \boldsymbol{\lambda}^{\mathsf{o}}_I ) \right) \; \text{ s.t. } \eqref{range_01}   \\
		& = \mathop{\mathrm{max}}_{\mathbf{x} \in \mathcal{X}} \;\;
		\mathsf{h}^{\mathsf{o}} \! \left(  
		\max_{\boldsymbol{\lambda}^{\mathsf{o}}_1} \; \tilde{\mathsf{f}}^{\mathsf{o}}_1 (\mathbf{x}, \boldsymbol{\lambda}^{\mathsf{o}}_1 ), 
		\ldots, 
		\max_{\boldsymbol{\lambda}^{\mathsf{o}}_I} \; \tilde{\mathsf{f}}^{\mathsf{o}}_I (\mathbf{x}, \boldsymbol{\lambda}^{\mathsf{o}}_I ) 
		\right)  \notag  \\
		& \quad  \text{ s.t. } \left(  
		\max_{\boldsymbol{\lambda}^{\mathsf{o}}_1} \; \tilde{\mathsf{f}}^{\mathsf{o}}_1 (\mathbf{x}, \boldsymbol{\lambda}^{\mathsf{o}}_1 ), 
		\ldots, 
		\max_{\boldsymbol{\lambda}^{\mathsf{o}}_I} \; \tilde{\mathsf{f}}^{\mathsf{o}}_I (\mathbf{x}, \boldsymbol{\lambda}^{\mathsf{o}}_I ) 
		\right) \in \mathcal{H}^{\mathsf{o}}  \label{constraint_H}  \\
		& = \mathop{\mathrm{max}}_{\mathbf{x} \in \mathcal{X}} \;\;
		\mathsf{h}^{\mathsf{o}} \! \left(  
		\max_{\boldsymbol{\lambda}^{\mathsf{o}}_1} \; \tilde{\mathsf{f}}^{\mathsf{o}}_1 (\mathbf{x}, \boldsymbol{\lambda}^{\mathsf{o}}_1 )
		\ldots, 
		\max_{\boldsymbol{\lambda}^{\mathsf{o}}_I} \; \tilde{\mathsf{f}}^{\mathsf{o}}_I (\mathbf{x}, \boldsymbol{\lambda}^{\mathsf{o}}_I ) 
		\right) . 
	\end{align}
\end{subequations}
The first equality follows from that $\mathsf{h}^{\mathsf{o}}(\cdot)$ is nondecreasing over $\mathcal{H}^{\mathsf{o}}$, and $\boldsymbol{\lambda}^{\mathsf{o}}_i$ are separable, so that we can move the max inside the function.
The second line follows from the relation in \eqref{87}, which implies that the constraint \eqref{constraint_H} can be dropped.

Next, we show that the feasible set of $\mathbf{x}$ satisfying \eqref{81b} is the same as that satisfying \eqref{29b} and \eqref{range_02}.
Let $(\mathbf{x}, \{\boldsymbol{\lambda}^{\mathsf{c}}_{k,j}\})$ be a point satisfying \eqref{29b} and \eqref{range_02}.
Then, $\mathbf{x}$ must satisfy \eqref{81b}.
This is because \eqref{81b} maximizes $\tilde{\mathsf{f}}^{\mathsf{c}}_{k,j} ( \mathbf{x}, \boldsymbol{\lambda}^{\mathsf{c}}_{k,j} )$ over $\boldsymbol{\lambda}^{\mathsf{c}}_{k,j}$.
From \eqref{88}, we know that the maximum still lies in the set $\mathcal{H}_k^{\mathsf{c}}$, over which the outer function $\mathsf{h}^{\mathsf{c}}_k(\cdot)$ is nondecreasing.

Conversely, if $\mathbf{x}$ is a feasible point in \eqref{81b}, based on \eqref{equivalence_02} and \eqref{88},
$( \mathbf{x}, \{ ( \boldsymbol{\lambda}^{\mathsf{c}}_{k,j})^{\star} \} )$ must be a feasible point in \eqref{29b} and \eqref{range_02}, where $(\boldsymbol{\lambda}^{\mathsf{c}}_{k,j})^{\star} = \arg \max_{\boldsymbol{\lambda}^{\mathsf{c}}_{k,j}} \tilde{\mathsf{f}}^{\mathsf{c}}_{k,j} ( \mathbf{x}, \boldsymbol{\lambda}^{\mathsf{c}}_{k,j} )$.
Therefore, the feasible set of $\mathbf{x}$ in the problem \eqref{equivalent Ray} is the same as that in the problem \eqref{81}.
Combined with the fact in \eqref{90}, this shows that the two problems are equivalent.

Secondly, we show the equivalence between problems \eqref{81} and \eqref{q}.  
According to the identities in \eqref{equivalence_01} and \eqref{equivalence_02}, problem \eqref{81} is equivalent to problem \eqref{q} since they share the same objective function and constraints.
Finally, combined with the fact that problem \eqref{equivalent Ray} is equivalent to problem \eqref{81}, problem \eqref{equivalent Ray} is shown to be equivalent to problem \eqref{q}.

\section{Proof of Lemma \ref{lem:linear_transform}}
\label{app:lemma_2}

Based on Lemma~\ref{lem:quadratic_transform}, we have the following relation:
\begin{align} \label{45}
	\mathsf{f}(\mathbf{x})
	\geq 
	2 \mathop{\mathfrak{Re}} \! \left( \boldsymbol{\lambda}^{\dagger} \mathbf{A} \mathbf{x} \right)
	- \boldsymbol{\lambda}^{\dagger} \mathbf{D}(\mathbf{x}) \boldsymbol{\lambda} .
\end{align}
The second term in \eqref{45} can be rewritten as
\begin{align} \label{93}
	\boldsymbol{\lambda}^{\dagger} \mathbf{D}(\mathbf{x}) \boldsymbol{\lambda}
	& = 
	\mathbf{x}^{\dagger}
	\left( \sum_{m} \varrho_m \left( \mathbf{B}_m^{\dagger}  \boldsymbol{\lambda} \right) \left( \mathbf{B}_m^{\dagger} \boldsymbol{\lambda} \right)^{\dagger} \right) \mathbf{x} + 
	\boldsymbol{\lambda}^{\dagger} \mathbf{C} \boldsymbol{\lambda}  \notag \\
	& \triangleq
	\mathbf{x}^{\dagger} \mathbf{M}(\boldsymbol{\lambda}) \mathbf{x}
	+ \boldsymbol{\lambda}^{\dagger} \mathbf{C} \boldsymbol{\lambda} .
\end{align}
Then, we eliminate the quadratic term in \eqref{93} by making use of the fact that for unit-modulus variables $\mathbf{x}$ and $\mathbf{z}$,
\begin{equation}
	\mathbf{x}^{\dagger} \left( \delta \mathbf{I} \right) \mathbf{x} 
	= \mathbf{z}^{\dagger} \left( \delta \mathbf{I} \right) \mathbf{z} 
	= \delta N .
\end{equation}
Specifically, we apply \cite[Eq. (26)]{7547360} to \eqref{93}, yielding
\begin{align} \label{MM}
	\mathbf{x}^{\dagger} \mathbf{M}(\boldsymbol{\lambda}) \mathbf{x}
	& \leq \mathbf{x}^{\dagger} \mathbf{L} \mathbf{x}
	+ \mathbf{z}^{\dagger} \left( \mathbf{L} - \mathbf{M}(\boldsymbol{\lambda})  \right) \mathbf{z}  \notag \\
	& + 2 \mathop{\mathfrak{Re}} \! \left\lbrace
	\mathbf{x}^{\dagger} 
	\left( \mathbf{M}(\boldsymbol{\lambda}) - \mathbf{L} \right) \mathbf{z} \right\rbrace,
\end{align}
where $\mathbf{L} \succeq \mathbf{M}(\boldsymbol{\lambda})$ and the equality is achieved at $\mathbf{z} = \mathbf{x}$.
Then, by replacing $\mathbf{L}$ with $\delta \mathbf{I}$, where $\delta$ is the trace of $\mathbf{M}(\boldsymbol{\lambda})$ so that $\delta \mathbf{I} \succeq \mathbf{M}(\boldsymbol{\lambda})$, 
and by combining with \eqref{45}, we obtain
\begin{align} \label{51}
	\mathsf{f}(\mathbf{x})
	& \geq
	2 \mathop{\mathfrak{Re}} \! \left\lbrace
	\mathbf{x}^{\dagger} \left(
	\left( \delta \mathbf{I} - \mathbf{M}(\boldsymbol{\lambda}) \right) \mathbf{z} + \mathbf{A}^{\dagger} \boldsymbol{\lambda}
	\right) \right\rbrace
	+ c \left( \mathbf{z}, \boldsymbol{\lambda} \right)  \notag  \\
	& \triangleq 
	\bar{\mathsf{f}}(\mathbf{x}, \mathbf{z}, \boldsymbol{\lambda}),
\end{align}
where $c \left( \mathbf{z}, \boldsymbol{\lambda} \right)$ is given in \eqref{constant_c}. 
The equality in the above is achieved when $\mathbf{z} = \mathbf{x}$ and $\boldsymbol{\lambda} = (\mathbf{D}(\mathbf{x}))^{-1} \mathbf{A} \mathbf{x}$.


\bibliographystyle{IEEEtran} 
\bibliography{reference}

\begin{thebibliography}{10}
\providecommand{\url}[1]{#1}
\csname url@samestyle\endcsname
\providecommand{\newblock}{\relax}
\providecommand{\bibinfo}[2]{#2}
\providecommand{\BIBentrySTDinterwordspacing}{\spaceskip=0pt\relax}
\providecommand{\BIBentryALTinterwordstretchfactor}{4}
\providecommand{\BIBentryALTinterwordspacing}{\spaceskip=\fontdimen2\font plus
\BIBentryALTinterwordstretchfactor\fontdimen3\font minus
  \fontdimen4\font\relax}
\providecommand{\BIBforeignlanguage}[2]{{%
\expandafter\ifx\csname l@#1\endcsname\relax
\typeout{** WARNING: IEEEtran.bst: No hyphenation pattern has been}%
\typeout{** loaded for the language `#1'. Using the pattern for}%
\typeout{** the default language instead.}%
\else
\language=\csname l@#1\endcsname
\fi
#2}}
\providecommand{\BIBdecl}{\relax}
\BIBdecl

\bibitem{Liu_GLOBECOM2024}
Y.~Liu and W.~Yu, ``{RIS}-assisted joint sensing and communications via
  fractionally constrained fractional programming,'' in \emph{Prof. IEEE Global
  Commun. Conf. (GLOBECOM)}, Cape Town, South Africa, Dec. 2024.

\bibitem{9737357}
F.~Liu, Y.~Cui, C.~Masouros, J.~Xu, T.~X. Han, Y.~C. Eldar, and S.~Buzzi,
  ``Integrated sensing and communications: Toward dual-functional wireless
  networks for {6G} and beyond,'' \emph{IEEE J. Sel. Areas Commun.}, vol.~40,
  no.~6, pp. 1728--1767, Jun. 2022.

\bibitem{9540344}
J.~A. Zhang, F.~Liu, C.~Masouros, R.~W. Heath, Z.~Feng, L.~Zheng, and
  A.~Petropulu, ``An overview of signal processing techniques for joint
  communication and radar sensing,'' \emph{IEEE J. Sel. Topics Signal
  Process.}, vol.~15, no.~6, pp. 1295--1315, Nov. 2021.

\bibitem{9606831}
Y.~Cui, F.~Liu, X.~Jing, and J.~Mu, ``Integrating sensing and communications
  for ubiquitous {IoT}: Applications, trends, and challenges,'' \emph{IEEE
  Netw.}, vol.~35, no.~5, pp. 158--167, Sep./Oct. 2021.

\bibitem{8741198}
C.~Huang, A.~Zappone, G.~C. Alexandropoulos, M.~Debbah, and C.~Yuen,
  ``Reconfigurable intelligent surfaces for energy efficiency in wireless
  communication,'' \emph{IEEE Trans. Wireless Commun.}, vol.~18, no.~8, pp.
  4157--4170, Aug. 2019.

\bibitem{8811733}
Q.~Wu and R.~Zhang, ``Intelligent reflecting surface enhanced wireless network
  via joint active and passive beamforming,'' \emph{IEEE Trans. Wireless
  Commun.}, vol.~18, no.~11, pp. 5394--5409, Nov. 2019.

\bibitem{10243495}
S.~P. Chepuri, N.~Shlezinger, F.~Liu, G.~C. Alexandropoulos, S.~Buzzi, and
  Y.~C. Eldar, ``Integrated sensing and communications with reconfigurable
  intelligent surfaces: From signal modeling to processing,'' \emph{IEEE Signal
  Process. Mag.}, vol.~40, no.~6, pp. 41--62, Sep. 2023.

\bibitem{8386661}
F.~Liu, L.~Zhou, C.~Masouros, A.~Li, W.~Luo, and A.~Petropulu, ``Toward
  dual-functional radar-communication systems: Optimal waveform design,''
  \emph{IEEE Trans. Signal Process.}, vol.~66, no.~16, pp. 4264--4279, Aug.
  2018.

\bibitem{9124713}
X.~Liu, T.~Huang, N.~Shlezinger, Y.~Liu, J.~Zhou, and Y.~C. Eldar, ``Joint
  transmit beamforming for multiuser {MIMO} communications and {MIMO} radar,''
  \emph{IEEE Trans. Signal Process.}, vol.~68, pp. 3929--3944, 2020.

\bibitem{9724202}
X.~Shao, C.~You, W.~Ma, X.~Chen, and R.~Zhang, ``Target sensing with
  intelligent reflecting surface: Architecture and performance,'' \emph{IEEE J.
  Sel. Areas Commun.}, vol.~40, no.~7, pp. 2070--2084, Jul. 2022.

\bibitem{9361184}
W.~Lu, Q.~Lin, N.~Song, Q.~Fang, X.~Hua, and B.~Deng, ``Target detection in
  intelligent reflecting surface aided distributed {MIMO} radar systems,''
  \emph{IEEE Sens. Lett.}, vol.~5, no.~3, pp. 1--4, Mar. 2021.

\bibitem{9454375}
S.~Buzzi, E.~Grossi, M.~Lops, and L.~Venturino, ``Radar target detection aided
  by reconfigurable intelligent surfaces,'' \emph{IEEE Signal Process. Lett.},
  vol.~28, pp. 1315--1319, 2021.

\bibitem{9732186}
------, ``Foundations of {MIMO} radar detection aided by reconfigurable
  intelligent surfaces,'' \emph{IEEE Trans. Signal Process.}, vol.~70, pp.
  1749--1763, 2022.

\bibitem{9769997}
R.~Liu, M.~Li, Y.~Liu, Q.~Wu, and Q.~Liu, ``Joint transmit waveform and passive
  beamforming design for {RIS}-aided {DFRC} systems,'' \emph{IEEE J. Sel.
  Topics Signal Process.}, vol.~16, no.~5, pp. 995--1010, Aug. 2022.

\bibitem{10364735}
R.~Liu, M.~Li, Q.~Liu, and A.~Lee~Swindlehurst, ``{SNR/CRB}-constrained joint
  beamforming and reflection designs for {RIS-ISAC} systems,'' \emph{IEEE
  Trans. Wireless Commun.}, vol.~23, no.~7, pp. 7456--7470, Jul. 2024.

\bibitem{9364358}
Z.-M. Jiang, M.~Rihan, P.~Zhang, L.~Huang, Q.~Deng, J.~Zhang, and E.~M.
  Mohamed, ``Intelligent reflecting surface aided dual-function radar and
  communication system,'' \emph{IEEE Syst. J.}, vol.~16, no.~1, pp. 475--486,
  Mar. 2022.

\bibitem{8999605}
F.~Liu, C.~Masouros, A.~P. Petropulu, H.~Griffiths, and L.~Hanzo, ``Joint radar
  and communication design: Applications, state-of-the-art, and the road
  ahead,'' \emph{IEEE Trans. Commun.}, vol.~68, no.~6, pp. 3834--3862, Jun.
  2020.

\bibitem{9618653}
M.~Temiz, E.~Alsusa, and M.~W. Baidas, ``A dual-function massive {MIMO} uplink
  {OFDM} communication and radar architecture,'' \emph{IEEE Trans. Cognit.
  Commun. Netw.}, vol.~8, no.~2, pp. 750--762, Jun. 2022.

\bibitem{9800940}
C.~Ouyang, Y.~Liu, and H.~Yang, ``Performance of downlink and uplink integrated
  sensing and communications ({ISAC}) systems,'' \emph{IEEE Wireless Commun.
  Lett.}, vol.~11, no.~9, pp. 1850--1854, Sept. 2022.

\bibitem{10608079}
B.~Zhao, C.~Ouyang, X.~Zhang, and Y.~Liu, ``Downlink and uplink {NOMA-ISAC}
  with signal alignment,'' \emph{IEEE Trans. Wireless Commun.}, vol.~23,
  no.~10, pp. 15\,322--15\,338, Oct. 2024.

\bibitem{10542219}
L.~Sun, Z.~Zhao, S.~Wang, Z.~Ding, and M.~Peng, ``On the study of
  non-orthogonal multiple access ({NOMA})-assisted integrated sensing and
  communication ({ISAC}),'' \emph{IEEE Trans. Commun.}, vol.~72, no.~11, pp.
  7278--7293, Nov. 2024.

\bibitem{10472418}
Q.~Qi, X.~Chen, C.~Zhong, C.~Yuen, and Z.~Zhang, ``Deep learning-based design
  of uplink integrated sensing and communication,'' \emph{IEEE Trans. Wireless
  Commun.}, vol.~23, no.~9, pp. 10\,639--10\,652, Sept. 2024.

\bibitem{10158711}
Z.~He, W.~Xu, H.~Shen, D.~W.~K. Ng, Y.~C. Eldar, and X.~You, ``Full-duplex
  communication for {ISAC}: Joint beamforming and power optimization,''
  \emph{IEEE J. Sel.d Areas Commun.}, vol.~41, no.~9, pp. 2920--2936, Sept.
  2023.

\bibitem{10557567}
R.~Li, L.~Wang, K.~Chen, L.~Xu, and A.~Fei, ``Full-duplex {NOMA}-enabled
  integrated sensing and communication: Joint transmit and receive beamforming
  optimization,'' \emph{IEEE Internet Things J.}, vol.~11, no.~16, pp.
  27\,015--27\,029, Aug. 2024.

\bibitem{10274660}
Y.~Guo, Y.~Liu, Q.~Wu, X.~Li, and Q.~Shi, ``Joint beamforming and power
  allocation for {RIS} aided full-duplex integrated sensing and uplink
  communication system,'' \emph{IEEE Trans. Wireless Commun.}, vol.~23, no.~5,
  pp. 4627--4642, May 2024.

\bibitem{9652071}
F.~Liu, Y.-F. Liu, A.~Li, C.~Masouros, and Y.~C. Eldar, ``{Cramér-Rao} bound
  optimization for joint radar-communication beamforming,'' \emph{IEEE Trans.
  Signal Process.}, vol.~70, pp. 240--253, 2022.

\bibitem{10138058}
X.~Song, J.~Xu, F.~Liu, T.~X. Han, and Y.~C. Eldar, ``Intelligent reflecting
  surface enabled sensing: {Cramér-Rao} bound optimization,'' \emph{IEEE
  Trans. Signal Process.}, vol.~71, pp. 2011--2026, 2023.

\bibitem{10440056}
X.~Song, X.~Qin, J.~Xu, and R.~Zhang, ``{Cramér-Rao} bound minimization for
  {IRS}-enabled multiuser integrated sensing and communications,'' \emph{IEEE
  Trans. Wireless Commun.}, vol.~23, no.~8, pp. 9714--9729, Aug. 2024.

\bibitem{9591331}
X.~Wang, Z.~Fei, J.~Huang, and H.~Yu, ``Joint waveform and discrete phase shift
  design for {RIS}-assisted integrated sensing and communication system under
  {Cramer-Rao} bound constraint,'' \emph{IEEE Trans. Veh. Technol.}, vol.~71,
  no.~1, pp. 1004--1009, Jan. 2022.

\bibitem{9416177}
X.~Wang, Z.~Fei, Z.~Zheng, and J.~Guo, ``Joint waveform design and passive
  beamforming for {RIS}-assisted dual-functional radar-communication system,''
  \emph{IEEE Trans. Veh. Technol.}, vol.~70, no.~5, pp. 5131--5136, May 2021.

\bibitem{6541985}
W.~Huleihel, J.~Tabrikian, and R.~Shavit, ``Optimal adaptive waveform design
  for cognitive {MIMO} radar,'' \emph{IEEE Trans. Signal Processing}, vol.~61,
  no.~20, pp. 5075--5089, 2013.

\bibitem{8314727}
K.~Shen and W.~Yu, ``Fractional programming for communication systems—{P}art
  {I}: Power control and beamforming,'' \emph{IEEE Trans. Signal Process.},
  vol.~66, no.~10, May 2018.

\bibitem{6847111}
A.~Alkhateeb, O.~El~Ayach, G.~Leus, and R.~W. Heath, ``Channel estimation and
  hybrid precoding for millimeter wave cellular systems,'' \emph{IEEE J. Sel.
  Topics Signal Process.}, vol.~8, no.~5, pp. 831--846, Oct. 2014.

\bibitem{9722893}
B.~Zheng, C.~You, W.~Mei, and R.~Zhang, ``A survey on channel estimation and
  practical passive beamforming design for intelligent reflecting surface aided
  wireless communications,'' \emph{IEEE Commun. Surv. Tutor.}, vol.~24, no.~2,
  pp. 1035--1071, Feb. 2022.

\bibitem{10663294}
Z.~Yu, H.~Ren, C.~Pan, G.~Zhou, R.~Wang, M.~Liu, and J.~Wang, ``Addressing the
  mutual interference in uplink isac receivers: A projection method,''
  \emph{IEEE Wireless Commun. Lett.}, vol.~13, no.~11, pp. 3109--3113, Nov.
  2024.

\bibitem{YuarXiv}
\BIBentryALTinterwordspacing
Z.~Yu, H.~Ren, C.~Pan, G.~Zhou, D.~Wang, C.~Yuen, and J.~Wang, ``A framework
  for uplink {ISAC} receiver designs: Performance analysis and algorithm
  development,'' \emph{arXiv}, Mar. 2025. [Online]. Available:
  \url{https://arxiv.org/abs/2503.02647}
\BIBentrySTDinterwordspacing

\bibitem{kay1993fundamentals}
S.~M. Kay, \emph{Fundamentals of Statistical Signal Processing: Estimation
  Theory}.\hskip 1em plus 0.5em minus 0.4em\relax Englewood Cliffs, NJ: PTR
  Prentice Hall, 1993.

\bibitem{9500437}
Y.~Liu, E.~Liu, R.~Wang, and Y.~Geng, ``Reconfigurable intelligent surface
  aided wireless localization,'' in \emph{Prof. IEEE Int. Conf. Commun.},
  Montreal, Canada, Jun. 2021, pp. 1--6.

\bibitem{van2002optimum}
H.~L. Van~Trees, \emph{Optimum Array Processing: Part IV of Detection,
  Estimation, and Modulation Theory}.\hskip 1em plus 0.5em minus 0.4em\relax
  Hoboken, NJ, USA: Wiley, 2002.

\bibitem{1523369}
J.~Dauwels, ``Computing {Bayesian Cramer-Rao} bounds,'' in \emph{Proc. IEEE
  Int. Symp. Inf. Theory (ISIT)}, Sept. 2005, pp. 425--429.

\bibitem{10584278}
C.~Xu and S.~Zhang, ``{MIMO} integrated sensing and communication exploiting
  prior information,'' \emph{IEEE J. Sel. Areas Commun.}, pp. 1--1, 2024.

\bibitem{1703855}
I.~Bekkerman and J.~Tabrikian, ``Target detection and localization using {MIMO}
  radars and sonars,'' \emph{IEEE Trans. Signal Process.}, vol.~54, no.~10, pp.
  3873--3883, 2006.

\bibitem{9931490}
X.~He and J.~Wang, ``{QCQP} with extra constant modulus constraints: Theory and
  application to {SINR} constrained mmwave hybrid beamforming,'' \emph{IEEE
  Trans. Signal Process.}, vol.~70, pp. 5237--5250, Oct. 2022.

\bibitem{boyd2011admm}
S.~Boyd, N.~Parikh, E.~Chu, B.~Peleato, and J.~Eckstein, ``Distributed
  optimization and statistical learning via the alternating direction method of
  multipliers,'' \emph{Found. Trends Mach. Learn.}, vol.~3, no.~1, pp. 1--122,
  Jan. 2011.

\bibitem{mosek2019}
{MOSEK ApS}, ``{MOSEK optimization toolbox for MATLAB},'' \emph{{U}ser’s
  Guide Reference Manual}, vol.~4, no.~1, 2019.

\bibitem{1055879}
R.~Miller and C.~Chang, ``A modified {Cramér-Rao} bound and its
  applications,'' \emph{IEEE Trans. Inf. Theory}, vol.~24, no.~3, pp. 398--400,
  May 1978.

\bibitem{7547360}
Y.~Sun, P.~Babu, and D.~P. Palomar, ``Majorization-minimization algorithms in
  signal processing, communications, and machine learning,'' \emph{IEEE Trans.
  Signal Process.}, vol.~65, no.~3, pp. 794--816, Feb. 2017.

\end{thebibliography}

\vfill

\end{document}